\documentclass{aa}
\usepackage{graphicx}
\usepackage{txfonts}
\usepackage{longtable,lscape}
\begin{document}
   \title{On the kinematic evolution of young local associations and 
          the Sco-Cen complex}

   \author{D. Fern\'andez\inst{1,2}\fnmsep\thanks{At present at Institut
           d'Estudis Espacials de Catalunya (IEEC), \mbox{c/. Gran Capit\`a 
	   2-4}, 08034 Barcelona, Spain.}
      \and F. Figueras\inst{1}
      \and J. Torra\inst{1}
          }

   \offprints{D. Fern\'andez, \\
   \email{david.fernandez@am.ub.es}}

   \institute{Departament d'Astronomia i Meteorologia, IEEC-Universitat de
              Barcelona, Av. Diagonal 647, 08028 Barcelona, Spain
          \and
              Observatori Astron\`omic del Montsec, Consorci del Montsec, 
              Pla\c ca Major 1, 25691 \`Ager (Lleida), Spain
             }

   \date{Received 26 April 2007 / Accepted 13 December 2007}

   \abstract
   {Over the last decade, several groups of young (mainly low-mass) stars
have been discovered in the solar neighbourhood (closer than $\sim$100 pc),
thanks to cross-correlation between X-ray, optical spectroscopy and
kinematic data. These young local associations -- including an important
fraction whose members are Hipparcos stars -- offer insights into the
star formation process in low-density environments, shed light on the
substellar domain, and could have played an important role in the recent
history of the local interstellar medium.}
   {To study the kinematic evolution of young local associations and their 
relation to other young stellar groups and structures in the local interstellar 
medium, thus casting new light on recent star formation processes in the solar 
neighbourhood.}
   {We compiled the data published in the literature for young local 
associations. Using a realistic Galactic potential we integrated the orbits for 
these associations and the Sco-Cen complex back in time.}
   {Combining these data with the spatial structure of the Local Bubble
and the spiral structure of the Galaxy, we propose a recent history of
star formation in the solar neighbourhood. We suggest that both the
Sco-Cen complex and young local associations originated as a result of the 
impact of the inner spiral arm shock wave against a giant molecular cloud. 
The core of the giant molecular cloud formed the Sco-Cen complex, and some small 
cloudlets in a halo around the giant molecular cloud formed young local 
associations several million years later. We also propose a supernova in young 
local associations a few million years ago as the most likely candidate to
have reheated the Local Bubble to its present temperature.}
   {}


   \maketitle
%

\section{Introduction}

The first moving groups of stars were discovered early on in the history
of Galactic dynamics. As early as 1869, R.A. Proctor published a work in
which he identified a group of stars around the Hyades open cluster that
was moving together through the Galaxy. He also found another 5 comoving
stars in the Ursa Major constellation. In the 1960s, O.J. Eggen suggested
the existence of a {\it Local Association} (see for example Eggen
\cite{Eggen61}, \cite{Eggen65a}, \cite{Eggen65b}) formed by a group of
young stars with approximately the same spatial velocity (also referred as
the {\it Pleiades moving group}). Eggen's Local Association included the
nearest bright B-type stars, stars in the Pleiades, the $\alpha$ Perseus
and IC2602 clusters, and the stars belonging to the Sco-Cen complex.
However, it was difficult to defend a unitary view of the group due to the
wide range of ages ($\la$100 Myr) and widespread spatial distribution
($\sim$300 pc) of its constituent stars.

At the same time as Eggen was studying his Local Association, the first
X-ray detectors were installed in rockets and launched into space.  This
led to the discovery of the diffuse soft X-ray background (SXRB). The
anticorrelation observed between the SXRB and the HI column density was
rapidly interpreted as evidence of a local cavity in the interstellar
medium (ISM) filled by an X-ray emitting plasma, which became known as the
{\it Local Bubble} (LB). The launch of the ROSAT satellite in 1990 allowed
the LB to be studied in more detail. The opening of the X-ray window in the
sky and observations of stars belonging to clusters with known ages also
led to the realisation that X-ray emission persists in young stars for a
period of the order of 100 Myr. An important fraction of these young,
X-ray emitting stars are T Tauri stars. These very young ($\la$10 Myr)
stars also exhibit excess IR emission (due to the presence of nearby
heated dust particles forming disks or envelopes) as well as UV line and
continuum emission produced by the accretion of the surrounding gas and
dust. When the stars reach the age of $\sim$10 Myr, both IR emissions and
optical activity decline considerably. X-ray activity then becomes the
basis for ascertaining youthfulness.

The ROSAT All-Sky Survey (RASS) detected more than 100,000 X-ray sources
at keV energies (Voges et al. \cite{Voges00}). Using a kinematic approach
and searching for groups of comoving stars allows us to determine which of
them are T Tauri stars. This greatly reduces the number of candidate stars
that must be observed spectroscopically to confirm their youthfulness.

In this way, several young stellar associations have been discovered within
100 pc of Earth during the last decade (see Jayawardhana
\cite{Jayawardhana00}, and Zuckerman \& Song \cite{Zuckerman04b} for a
recent review). The stars belonging to these young local associations
(YLA) have ages ranging from a few million to several tens of millions of
years. Due to their proximity to us, these stars are spread over a large
area of the sky (up to several hundred square degrees) making it
difficult to identify the associations as single entities. Furthermore,
they are far away from molecular clouds or star forming regions (SFR).
This is why they remained unnoticed until recently.

These young stars offer insights into the star formation process in
low-density environments, which is different from the dominant mechanism
observed in denser SFR. The discovery of these YLA has also contributed to
substellar astrophysics, since dozens of dwarfs have been identified in
them. For instance, it has been confirmed that isolated brown dwarfs can
form in these low-density environments (see for example Webb et al.
\cite{Webb99}; Lowrance et al. \cite{Lowrance99}; Chauvin et al.
\cite{Chauvin03}). YLA also provide important clues regarding recent star
formation in the solar neighbourhood (since they contain the youngest
stars near the Sun) and its effect on the local ISM.

In Sect. \ref{sect.LB+YLA} of this paper we review current understanding
of the LB and the Sco-Cen complex and present published data for the nearly
300 stars belonging to YLA. In Sect. \ref{sect.orbits} we present our
method for orbit integration, which makes use of a Galactic potential that
includes the general axisymmetric potential and the perturbations due to
the Galaxy's spiral structure and the central bar. This leads us to study,
in Sect. \ref{sect.origin}, the origin and evolution of the local
structures and, in Sect. \ref{sect.sce}, to propose a scenario for recent
star formation in the solar neighbourhood. This scenario explains the
origin of the Sco-Cen complex and YLA through the collision of a parent
giant molecular cloud (GMC) with the Sagittarius-Carina spiral arm.  
Finally, the conclusions of our work are summarised in Sect.  
\ref{sect.conc}.

%

\section{The nearest solar neighbourhood}
\label{sect.LB+YLA}

In this section we consider three important structures: the Local Bubble
(its spatial structure is presented); the Sco-Cen(-Lupus-Crux) complex
(spatial structure, kinematics and age are presented); and the recently 
discovered YLA (for which we have produced a complete compendium of their 
members and mean properties).

\subsection{The Local Bubble}\label{subsec.LB}

Locally, within the nearest 100 pc, the ISM is dominated by the LB.  The
displacement model (Snowden et al. \cite{Snowden90}, \cite{Snowden98}) for
this structure assumes that the irregular local HI cavity is filled by an
X-ray-emitting plasma, with an emission temperature of $\sim10^6$ K and a
density, $n_e$, of $\sim 0.005$ cm$^{-3}$. Using the distance to the MBM
12 molecular cloud as a distance scale, Snowden et al.  
(\cite{Snowden98}) derived an extension for the LB of 40 to 130 pc, it
being larger at higher Galactic latitudes and smaller nearer the equator.
Nevertheless, there is no agreement in the literature concerning the
distance to MBM 12; values range from 60 to 300 pc (see for example Hearty
et al. \cite{Hearty00a}, \cite{Hearty00b}; Luhman \cite{Luhman01};
Andersson et al. \cite{Andersson02}). This could explain why the extension
of the LB calculated from SXRB observations does not agree (by a few tens of
parsecs) with the extension of the local cavity derived from HI
observations (Paresce \cite{Paresce84}; Welsh et al. \cite{Welsh94}).

\begin{figure}
\centering
\resizebox{7cm}{!}{\includegraphics{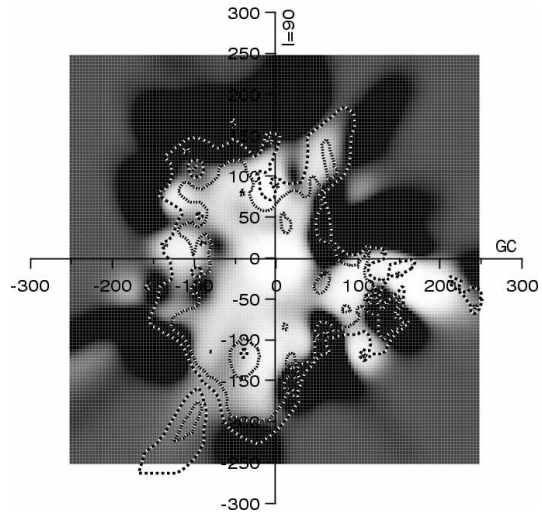}}
\caption{3D volumic density of the Local Bubble from the iso equivalent 
         width contours for $W$ = 20 mÅ and 50 mÅ (from 
         Lallement et al. \cite{Lallement03}).}
\label{fig.LB_Lallement}
\end{figure}

Sfeir at al. (\cite{Sfeir99}) and Lallement et al. (\cite{Lallement03})  
obtained the contours of the LB from NaI absorption measurements taken towards
a selected set of stellar targets with Hipparcos parallaxes of up to 350
pc from the Sun. These observations allowed the authors to draw maps of
the neutral gas distribution in the local ISM and, in particular, to trace
the contours and extension of the LB with an estimated precision of
$\approx\pm$20 pc in most directions. Lallement et al.
(\cite{Lallement03}) concluded that for heliocentric distances of about
70-90 pc, the equivalent width of the NaI D-line doublet increases from
$<$5 m\AA \, to $>$30 m\AA. They interpret this sudden increase as a
neutral gas boundary of the local cavity, i.e. the LB limit. This
boundary (see Fig.  \ref{fig.LB_Lallement}) has some {\it interstellar
tunnels} that connect the LB with other cavities, such as Loop I and the
supershell GSH238+00+09. In the meridian plane, the structure is clearly
observed to tilt by about 20\degr (see Fig. 5 in Lallement et al.
\cite{Lallement03}). This tilt is probably the result of {\it squeezing}
of the LB by the surrounding structures belonging to the Gould Belt, which
form a local chimney perpendicular to the plane of the Belt. (Sfeir et al.
\cite{Sfeir99}; Welsh et al. \cite{Welsh99}).

Some LB formation models have been proposed (see for example Cox
\cite{Cox98}), most of which involve several recent supernovae (SNe). The
Sco-Cen complex is the most likely candidate to have sheltered these SNe
(Ma\'{\i}z-Apell\'aniz \cite{MaizApellaniz01}; Fuchs et al. 
\cite{Fuchs06}).

\subsection{The Sco-Cen complex}\label{subsect.ScoCen}

The Sco-Cen complex dominates the fourth Galactic quadrant (de Zeeuw et
al. \cite{Zeeuw99}; hereafter Z99). It is a region of recent star formation 
and contains an important fraction of the most massive stars in the solar
neighbourhood. In the 1960s the complex was split into three components
(Blaauw \cite{Blaauw60}): Upper Scorpius (US), Upper Centaurus Lupus (UCL)
and Lower Centaurus Crux (LCC). The ISM related to this complex was
studied by de Geus (\cite{Geus92}), who found that there is not much
interstellar material associated with UCL and LCC, whereas filamentous
structures connect US with the cloud complex in Ophiuchus. The most widely
accepted ages for the components of the complex were derived by de Geus et
al.  (\cite{Geus89}) from isochrones in the HR diagram: $\sim$5-6 Myr for
US, $\sim$14-15 Myr for UCL and $\sim$11-12 Myr for LCC. Blaauw
(\cite{Blaauw64}, \cite{Blaauw91}) suggested that this age progression was
the result of a sequence of star formation events in the GMC that formed
all of the Sco-Cen region. These {\it classical} ages have recently been
questioned by studies of the low-mass component of the complex. The ages
obtained from such studies are in the range of 8-10 Myr for US and 16-20
Myr for UCL and LCC (Sartori et al. \cite{Sartori01}; Mamajek et al.
\cite{Mamajek02}; Sartori et al. \cite{Sartori03}, hereafter S03).  
However, Preibisch et al. (\cite{Preibisch02}) once again obtained an age
of $\sim$5 Myr for US from the HR diagram of the stars belonging to this
association with masses between 0.1 and 20 M$_{\sun}$.

Z99, searching in the Hipparcos catalogue, found 120 members of US, 221 of UCL 
and 180 of LCC, with mean trigonometric distances (corrected for systematic 
effects) of $145\pm2$, $140\pm2$ and $118\pm2$ pc, respectively (see Table 
\ref{tab.Assoc}). With regard to kinematics, all three associations have large 
velocity components in the direction away from the Sun, classically associated 
with the expansion motion of the Gould Belt (see for example P\"oppel 
\cite{Poppel97}, but also see Torra et al. \cite{Torra00}). Z99 provide mean 
proper motions from the Hipparcos catalogue and radial velocities from the 
Hipparcos Input catalogue for the stars identified as members of US, UCL and 
LCC. Madsen et al. (\cite{Madsen02}; hereafter M02) obtain astrometric radial 
velocities for the members of the Sco-Cen associations identified by Z99, 
assuming that all members share the same velocity vector. S03 compile radial 
velocities for the Sco-Cen members and derive from them the mean spatial motion, 
using Hipparcos parallaxes and proper motions. The mean spatial motions and 
standard deviations obtained by these authors are shown in Table 
\ref{tab.Assoc}. We can see that the results are quite similar, except for the 
case of US, where there is a clear discrepancy in the $U$ value. As stated by 
M02, US is close to the limit of their method (larger distance, smaller angular 
size and smaller number of member stars, in comparison with LCC and UCL); thus, 
significant biases may be introduced into the solution they obtained for this 
association. Therefore, hereafter in this paper the Sco-Cen kinematics from S03 
have been used to compute the back-tracing motion of the three associations. 
We have only used Madsen's data in Table \ref{tab.DistAss}.

\subsection{Young local associations}
\label{subsec.YLA}

\begin{table*}
   \caption{Mean spatial coordinates and heliocentric velocity components 
            of the young local associations and the Sco-Cen complex (in 
            the latter case, data from Z99, M02 and S03). 
            In brackets, the standard deviation of the sampling 
            distribution. $N$ is the number of known members in each 
	    association ($N_\mathrm{r}$ with distance determination and 
	    $N_\mathrm{k}$ with complete kinematic data).}
   \label{tab.Assoc}
\centering
\begin{tabular}{lrrrrrrrrrrr}
\hline
\hline
Association     & $\overline{\xi^\prime}$ &
                  $\overline{\eta^\prime}$ &
                  $\overline{\zeta^\prime}$
                & $\overline{r}$
                & $\overline{U}$ & $\overline{V}$ & $\overline{W}$
                & Age & $N$ & $N_\mathrm{r}$ & $N_\mathrm{k}$ \\
                & (pc) & (pc) & (pc) & (pc) 
		& (km s$^{-1}$) & (km s$^{-1}$) & (km s$^{-1}$) & (Myr)
		& & & \\
\hline
TW Hya          & $-21_{(22)}$
                & $-53_{(23)}$
                & $21_{(\;\;7)}$
                &  $63_{(30)}$
                &  $-9.7_{(4.1)}$ & $-17.1_{(3.1)}$ &  $-4.8_{(3.7)}$
                &  8 & 39 & 19 & 17 \\
Tuc-Hor/GAYA    & $-12_{(22)}$
                & $-24_{(11)}$
                & $-34_{(\;\;8)}$
                &  $49_{(\;\;8)}$
                & $-10.1_{(2.4)}$ & $-20.7_{(2.3)}$ &  $-2.5_{(3.8)}$
                &  20 & 52 & 50 & 44 \\
$\beta$ Pic-Cap & $-9_{(27)}$
                & $-5_{(14)}$
                & $-15_{(10)}$
                &  $35_{(11)}$
                & $-10.8_{(3.4)}$ & $-15.9_{(1.2)}$ &  $-9.8_{(2.5)}$
                &  12 & 33 & 24 & 24 \\
$\epsilon$ Cha  & $-47_{(\;\;8)}$
                & $-80_{(14)}$
                & $-25_{(\;\;5)}$
                &  $96_{(17)}$
                &  $-8.6_{(3.6)}$ & $-18.6_{(0.8)}$ &  $-9.3_{(1.7)}$
                &  10 & 16 & 6 & 5 \\
$\eta$ Cha      & $-33_{(\;\;2)}$
                & $-80_{(\;\;5)}$
                & $-34_{(\;\;2)}$
                &  $93_{(\;\;6)}$
                & $-12.2_{(0.0)}$ & $-18.1_{(0.9)}$ & $-10.1_{(0.5)}$
                &  10 & 18 & 3 & 2 \\
HD 141569       & $-77_{(\;\;3)}$
                & $10_{(\;\;8)}$
                & $64_{(\;\;8)}$
                & $101_{(\;\;8)}$
                &  $-5.4_{(1.5)}$ & $-15.6_{(2.6)}$ &  $-4.4_{(0.8)}$
                &  5 & 5 & 3 & 2 \\
Ext. R CrA      & $-97_{(44)}$
                & $-1_{(\;\;4)}$
                & $-30_{(16)}$
                & $102_{(47)}$
                &  $-0.1_{(6.4)}$ & $-14.8_{(1.4)}$ & $-10.1_{(3.3)}$
                &  13 & 59 & 7 & 4 \\
AB Dor          & $6_{(21)}$
                & $4_{(17)}$
                & $-12_{(16)}$
                & $32_{(13)}$
                &  $-7.4_{(3.2)}$ & $-27.4_{(3.2)}$ & $-12.9_{(6.4)}$
                &  30-150 & 40 & 36 & 35 \\
\hline
US \hspace{0.53cm}Z99 & $-141_{(34)}$ & $-22_{(11)}$ & $50_{(16)}$
                     & $145_{(\;\;2)}$
                     & & &
                     & 5-6 & 120 &  &  \\
\hspace{1.0cm}M02$^1$ & $-138_{(27)}$ & $-22_{(10)}$ & $49_{(12)}$
                     & $149_{(28)}$
                     & $-0.9\;\;\;\;\,$ & $-16.9\;\;\;\;\,$ & $-5.2\;\;\;\;\,$
                     & & 120 &  &  \\
\hspace{1.0cm}S03    & & & &
                     & $-6.7_{(5.9)}$ & $-16.0_{(3.5)}$ & $-8.0_{(2.7)}$
                     & 8-10 & 155 &  &  \\
\hline
UCL \hspace{0.3cm}Z99 & $-122_{(30)}$ & $-69_{(26)}$ & $32_{(16)}$
                     & $140_{(\;\;2)}$
                     & & &
                     &  14-15 & 221 &  &  \\
\hspace{1.0cm}M02$^1$ & $-121_{(26)}$ & $-68_{(21)}$ & $32_{(15)}$
                     & $145_{(24)}$
                     & $-7.9\;\;\;\;\,$ & $-19.0\;\;\;\;\,$ & $-5.7\;\;\;\;\,$
                     & & 218 &  &  \\
\hspace{1.0cm}S03    & & & &
                     & $-6.8_{(4.6)}$ & $-19.3_{(4.7)}$ & $-5.7_{(2.5)}$
                     & 16-20 & 262 &  &  \\
\hline
LCC \hspace{0.3cm}Z99 & $-62_{(18)}$ & $-102_{(24)}$ & $14_{(16)}$
                     & $118_{(\;\;2)}$
                     & & &
                     &  11-12 & 180 &  &  \\
\hspace{1.0cm}M02$^1$ & $-61_{(14)}$ & $-100_{(15)}$ & $14_{(15)}$
                     & $120_{(18)}$
                     & $-11.8\;\;\;\;\,$ & $-15.0\;\;\;\;\,$ & $-6.7\;\;\;\;\,$
                     & & 179 &  &  \\
\hspace{1.0cm}S03    & & & &
                     & $-8.2_{(5.1)}$ & $-18.6_{(7.3)}$ & $-6.4_{(2.6)}$
                     & 16-20 & 192 &  &  \\
\hline
\multicolumn{12}{l}{\tiny $^1$ M02 derived an internal velocity dispersion among individual
stars of 1.33 km s$^{-1}$ for US, 1.23 km s$^{-1}$ for UCL and 1.13 km s$^{-1}$ 
for LCC.} \\
\end{tabular}
\end{table*}

\begin{table*}
   \caption{Estimated ages for the YLA published in the literature.}
   \label{tab.AssocAges}
\centering
\begin{tabular}{llll}
\hline
\hline
Association     & Estimated age       & Method             & Reference \\
\hline
TW Hya          & 10$^{+10}_{-5}$ Myr & Spectroscopy + HR diagram (BVI) & Soderblom et al. (\cite{Soderblom98}) \\
                & $\sim$ 8 Myr        & HR diagram (JHK)   & Webb et al. (\cite{Webb99}) \\
                & 5-15 Myr            & HD diagram (IR)    & Weintraub et al. (\cite{Weintraub00}) \\
	        & 8.3 Myr             & Expansion age      & Makarov \& Fabricius (\cite{Makarov01}) \\
	        & 4.7 $\pm$ 0.6 Myr   & Expansion age      & Makarov et al. (\cite{Makarov05}) \\
	        & 20$^{+25}_{-7}$ Myr & Expansion age      & Mamajek (\cite{Mamajek05}) \\
	        & 8.3 $\pm$ 0.8 Myr   & Dynamical age      & de la Reza et al. (\cite{delaReza06}) \\
	        & 10$^{+10}_{-7}$ Myr & Spec. + HR diagram (VIJK) + Li + H$\alpha$ & Barrado y Navascu\'es (\cite{Barrado06}) \\
\hline
Tuc-Hor/GAYA    & $\sim$40 Myr        & H$\alpha$ emission & Zuckerman \& Webb (\cite{Zuckerman00}) \\
                & 10-30 Myr           & X-ray emission     & Stelzer \& Ne\"uhauser (\cite{Stelzer00}) \\
	        & $\sim$30 Myr        & Kinematic + HR diagram (BV)      & Torres et al. (\cite{Torres00}) \\
	        & 10-40 Myr           & HR diagram (VRI)   & Zuckerman et al. (\cite{Zuckerman01b}) \\
		& $\sim$20 Myr        & Kinematic age      & Torres et al. (\cite{Torres01}) \\
		& 27 Myr              & Dynamical age      & Makarov (\cite{Makarov07}) \\
\hline
$\beta$ Pic-Cap & 20 $\pm$ 10 Myr     & HR diagram (BVI)   & Barrado y Navascu\'es (\cite{Barrado99}) \\
                & 12$^{+8}_{-4}$ Myr  & HR diagram (VI) + Li & Zuckerman et al. (\cite{Zuckerman01a}) \\
	        & 11.2 $\pm$ 0.3 Myr  & Dynamical age      & Ortega et al. (\cite{Ortega02}) \\
		& 12 Myr              & Dynamical age      & Song et al. (\cite{Song03}) \\
		& 22 $\pm$ 12 Myr     & Dynamical age      & Makarov (\cite{Makarov07}) \\ 
\hline
$\epsilon$ Cha  & 5-15 Myr            & HR diagram (BV)    & Terranegra et al. (\cite{Terranegra99}) \\
                & 3-5 Myr             & HR diagram (VI)    & Feigelson et al. (\cite{Feigelson03}) \\
                & 7 Myr               & Dynamical age      & Jilinski et al. (\cite{Jilinski05}) \\
\hline
$\eta$ Cha      & 2-18 Myr            & HR diagram (R) + Li & Mamajek et al. (\cite{Mamajek99}) \\
                & 10-15 Myr           & Dynamical age      & Mamajek et al. (\cite{Mamajek00}) \\
		& 5 $\pm$ 4 Myr       & HR diagram (VRI)   & Lawson et al. (\cite{Lawson01}) \\
		& 6$^{+2}_{-1}$ Myr   & HR diagram (JHK)   & Luhman \& Steeghs (\cite{Luhman04}) \\
		& 7 Myr               & Dynamical age      & Jilinski et al. (\cite{Jilinski05}) \\
\hline
HD 141569       & 5 $\pm$ 3 Myr       & X-ray + Li + HR diagram (JHK) & Weinberger et al. (\cite{Weinberger00}) \\
                & 4.7 $\pm$ 0.3 Myr   & Spectroscopy + Kurucz's (\cite{Kurucz93}) models & Mer\'{\i}n et al. (\cite{Merin04}) \\
\hline
Ext. R CrA      & 10-15 Myr           & HR diagram (BVRI-JHK) & Neuha\"user et al. (\cite{Neuhauser00}) \\
                & $\sim$15 Myr        & Dynamical age      & Quast et al. (\cite{Quast01}) \\
\hline
AB Dor          & $\sim$ 50 Myr       & H$\alpha$ emission & Zuckerman et al. (\cite{Zuckerman04a}) \\
                & 75-150 Myr          & HR diagram (VK)    & Luhman et al. (\cite{Luhman05}) \\
		& 30-50 Myr and 80-120 Myr (2 subgroups) & HR diagram (VI) + Li & L\'opez-Santiago et al. (\cite{Lopez06}) \\
		& 38 Myr              & Dynamical age      & Makarov (\cite{Makarov07}) \\ 
		& 118 $\pm$ 20 Myr    & Dynamical age      & Ortega et al. (\cite{Ortega07}) \\
\hline
\end{tabular}
\end{table*}

A decade ago, very few PMS stars had been identified less than 100 pc from
the Sun. Nearly all the youngest stars ($\la$30 Myr) studied then were
located more than 140 pc away in the molecular clouds of Taurus,
Chamaeleon, Lupus, Sco-Cen and R CrA (all of them regions of recent star
formation). The cross-correlation of the Hipparcos and ROSAT catalogues
suddenly changed this; a few stars were identified as very young (from
their X-ray emission and lithium content) but closer than 100 pc, where
there are no molecular clouds with SFR (see for example Neuh\"auser \&
Brandner \cite{Neuhauser98}). Two explanations for the existence of these
young stars far away from SFR were proposed. Sterzik \& Durisen
(\cite{Sterzik95}) suggested that the stars were formed in molecular
clouds and later ejected as high-velocity stars during the decay of young
multiple star systems. Feigelson (\cite{Feigelson96}) suggested that the
stars were formed inside small molecular clouds (or {\it cloudlets}), which
later dispersed among the ISM and therefore can no longer be detected.

These young nearby stars were grouped into clusters, associations and
moving groups, each with a few dozen members. Different approaches were
used in each YLA discovery, but most of them made use of Hipparcos proper
motions, X-ray emission (as a youth indicator), infrared emission (youth
indicator) and ground-based spectroscopy (H$\alpha$ lines: youth
indicator, Li lines: age estimation, radial velocities) and photometry
(ages). We have compiled all the published YLA data in Appendix
\ref{sec.app}. Table \ref{tab.Assoc} shows the mean spatial and kinematic
properties, ages and number of members for each YLA. The adopted age shown in 
Table \ref{tab.Assoc} is that assigned for back-tracing the association
orbits in the next section (see Table \ref{tab.AssocAges} for an exhaustive 
list of estimated values found in the literature). More information and
tables containing all the data (in ASCII format) are available on a
webpage we have
published\footnote{http://www.am.ub.es/$\sim$dfernand/YLA}.

Different nomenclature is used by the authors to define each YLA. The
nature of the $\eta$ Cha cluster is clear: the stars are compacted in a
region a very few pc in diameter. For the other groups, authors used the
terms {\it association}, {\it moving group} or even {\it system} (for HD
141569, with only 5 known members). In this paper we have respected the
denomination for each YLA used in the literature.  However, the difference
between the terms {\it association} and {\it moving group} applied to the
different entities is not clear, since their spatial dispersions and
kinematics are very similar (and if slight differences are apparent, they
do not correspond to the classifications {\it association} or {\it moving
group}; see Table \ref{tab.Assoc}).

\begin{figure*}
\centering
\resizebox{8.5cm}{!}{\includegraphics{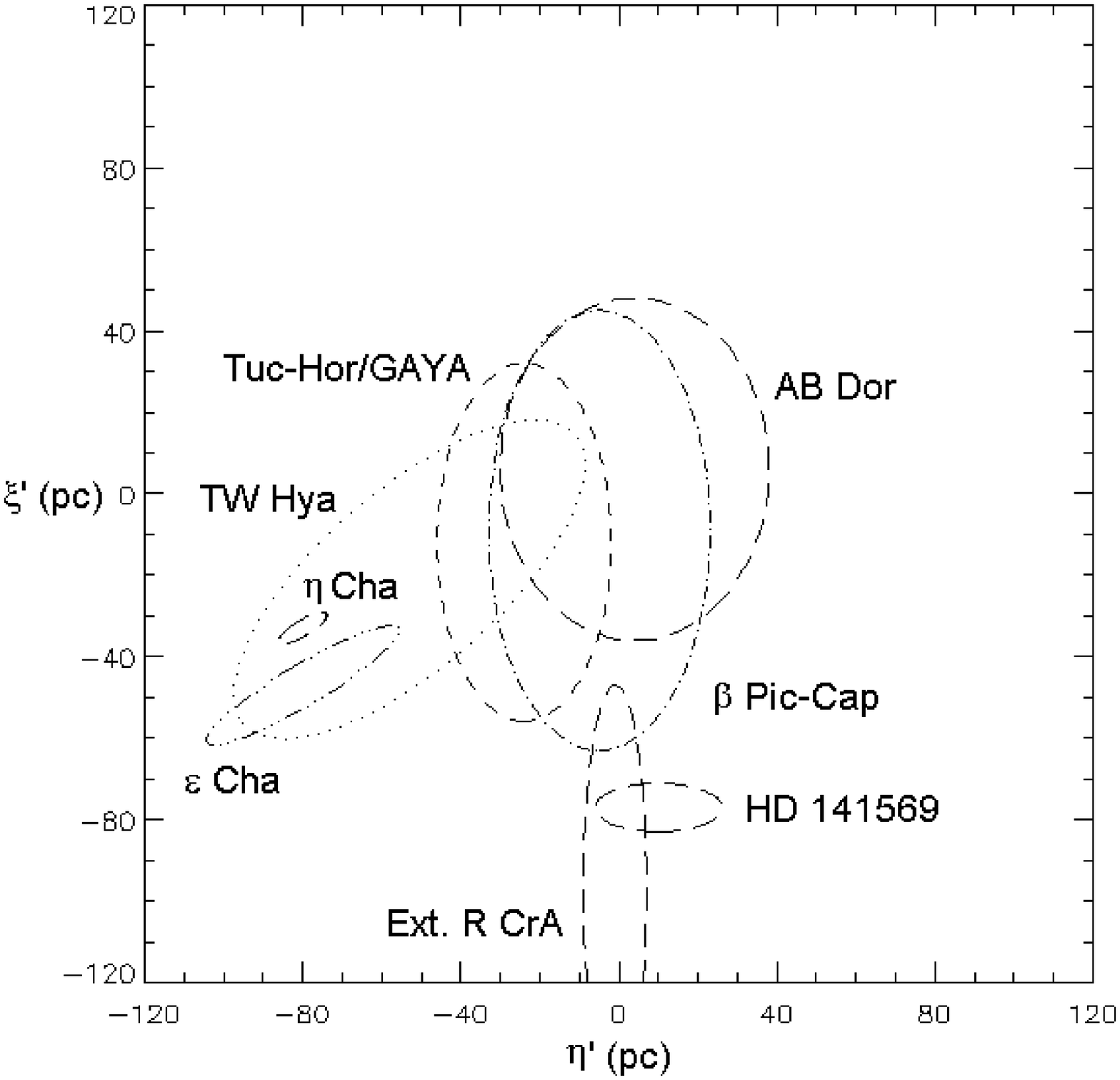}}
\hfill
\leavevmode
\resizebox{8.5cm}{!}{\includegraphics{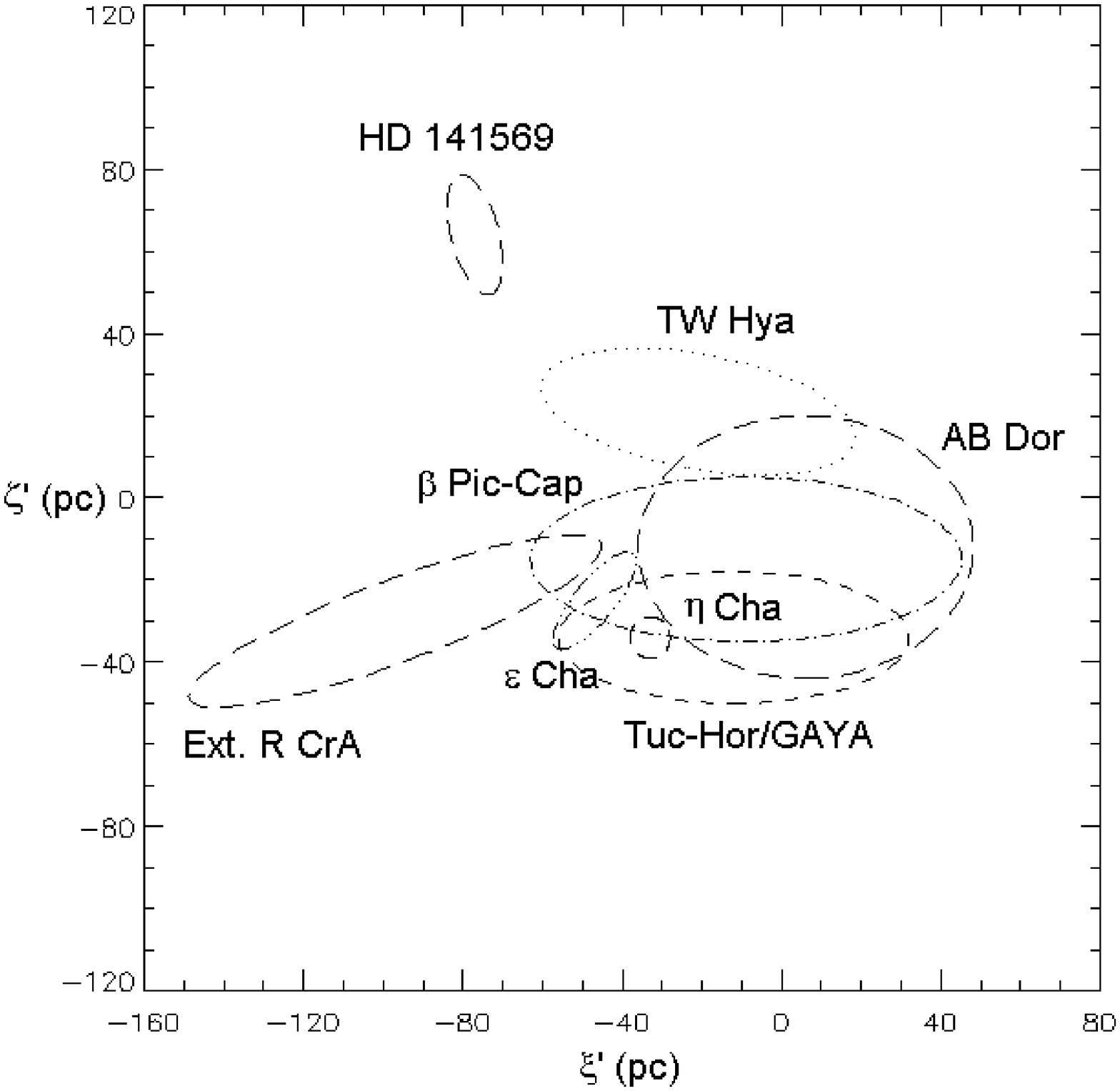}}
\hfill
\caption{Locations of the young local associations projected onto the 
         Galactic (left) and meridian (right) planes. The size of the 
	 ellipses represents the projected dimension of the associations (more 
	 than 90\% of the stars inside the plotted area). $\xi'$ is pointing to
         the Galactic anti-centre, $\eta'$ is the direction of Galactic rotation 
	 and $\zeta'$ is pointing to the north Galactic pole.
	 }
\label{fig.YLA}
\end{figure*}

Except for the case of AB Dor, all YLA studied here are younger than $\sim$30
Myr (even considering the large uncertainties in the age estimations published
in the literature; see Table \ref{tab.AssocAges}). The Tuc-Hor/GAYA association has 
an estimated age of about 20 Myr, whereas the other YLA (TW Hya, $\beta$ Pic-Cap, 
$\epsilon$ Cha, $\eta$ Cha, HD 141569 and Ext. R CrA) are younger than $\sim$15 Myr 
and of similar ages to (or younger than) the Sco-Cen associations. (The AB Dor moving 
group is clearly older and will not be used in our analysis; see, however, 
L\'opez-Santiago et al. \cite{Lopez06} and Makarov \cite{Makarov07}.)

At present, all YLA members in our compendium are spread over a region of
about 120 x 130 x 140 pc, mainly concentrated in the fourth Galactic
quadrant, and mostly in the southern celestial hemisphere (see Fig.
\ref{fig.YLA}). The spatial distributions within each association reach a
few tens of pc in the Galactic plane, and less than 20 pc in the
perpendicular direction (see Table \ref{tab.Assoc}).

The heliocentric velocity components of all the associations are very
similar, $(U,V,W) \sim \lbrack -$(9-12), $-$(16-21), $-$(3-10)$\rbrack$ km
s$^{-1}$ (except for the extended R CrA association and, maybe, the HD
141569 system). To obtain these mean values we used all the stars for
which complete kinematic data are available. We checked that rejecting
those stars with the larger residuals does not modify the mean values by
more than 1.0 km s$^{-1}$, though the standard deviations are clearly
reduced (in most cases less than 2.0 km s$^{-1}$ for all the components).
The only exception is the extended R CrA association. It contains only 5
stars with complete kinematic data and there are large uncertainties in
their radial velocities. This results in large standard deviations in the
velocity components, especially in $U$, and prevents us from using this
association in our analysis.

\begin{figure*}
\centering
\resizebox{6cm}{!}{\includegraphics{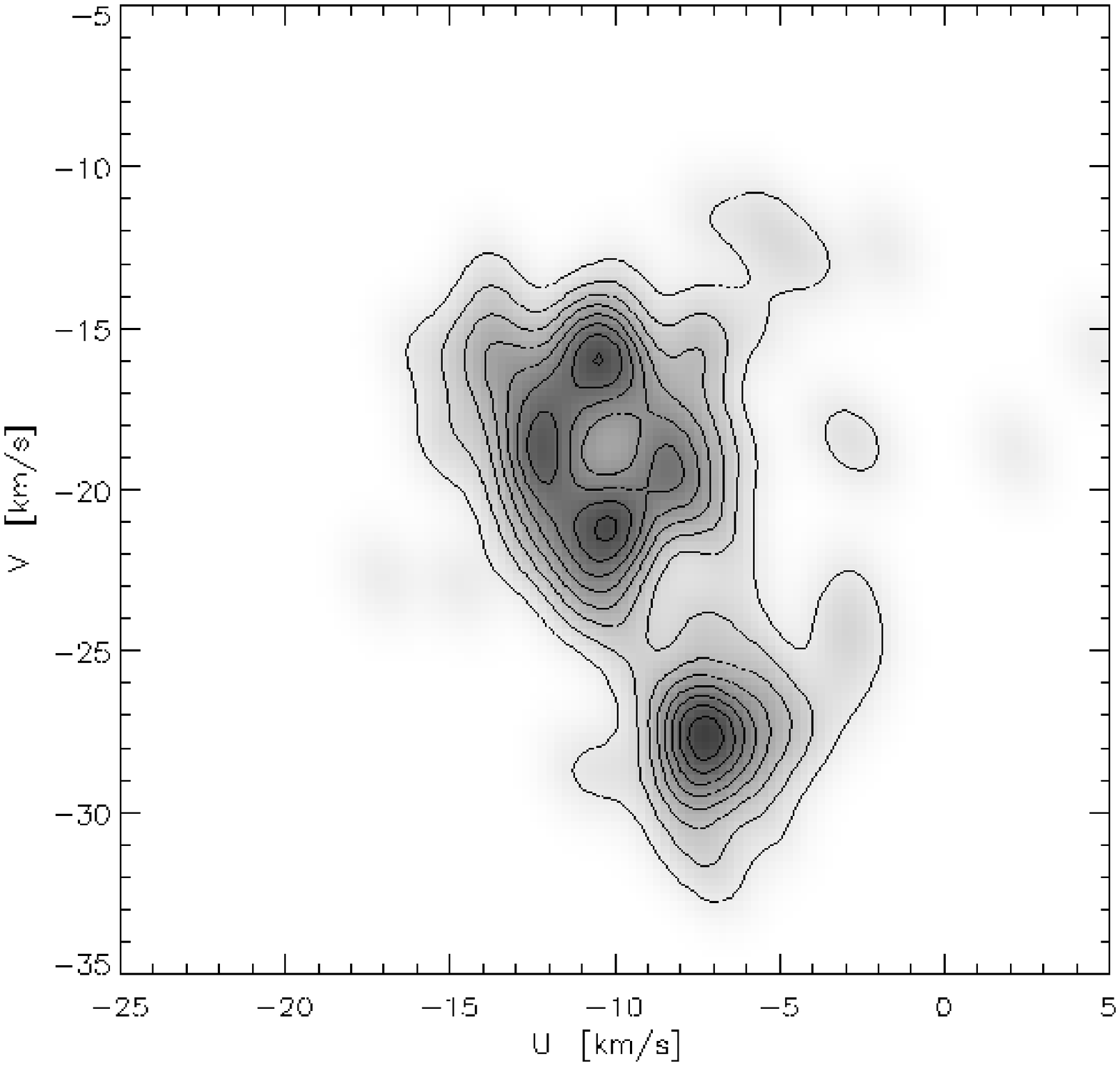}}
\hfill
\leavevmode
\resizebox{6cm}{!}{\includegraphics{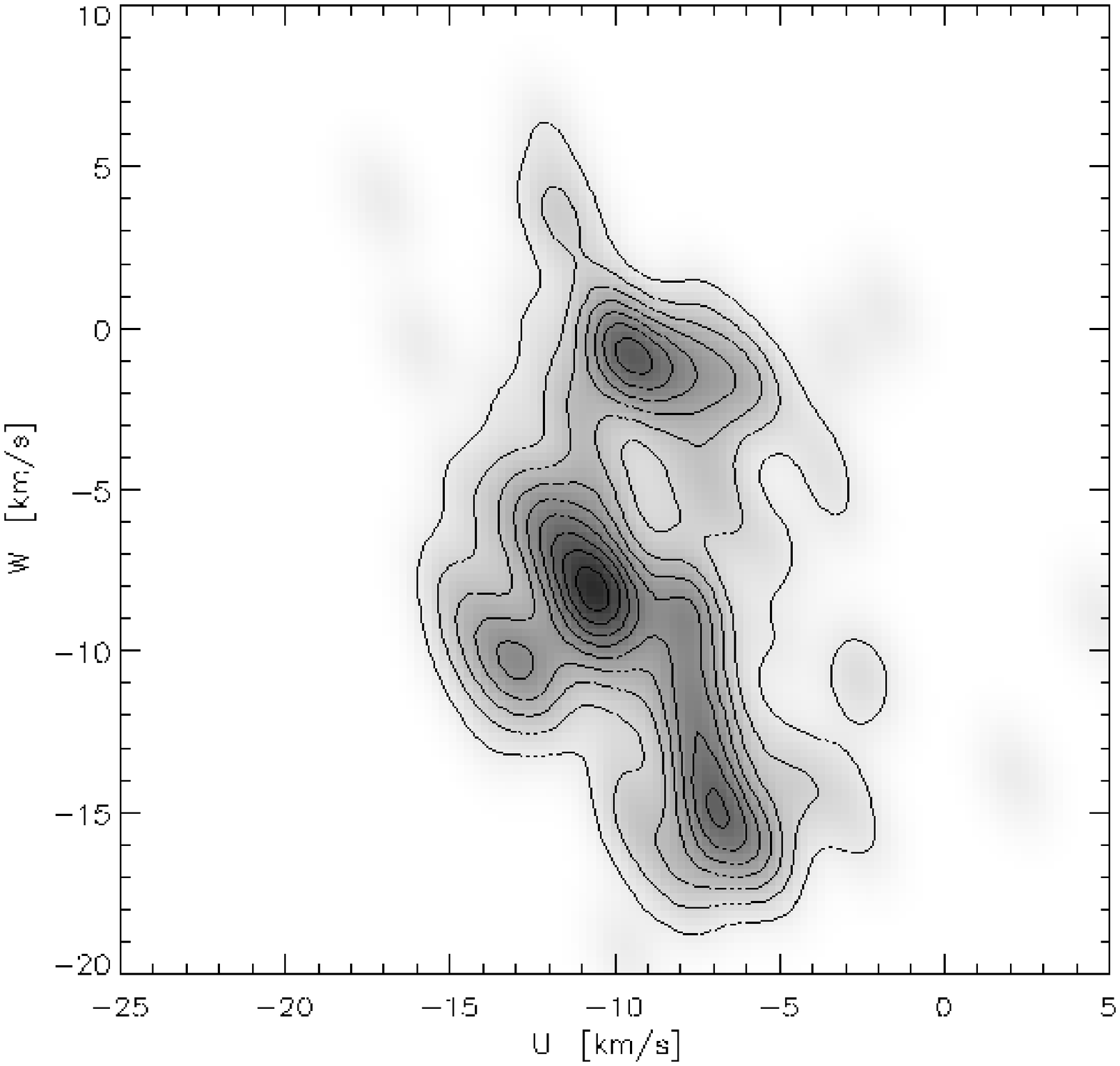}}
\hfill
\leavevmode
\resizebox{6cm}{!}{\includegraphics{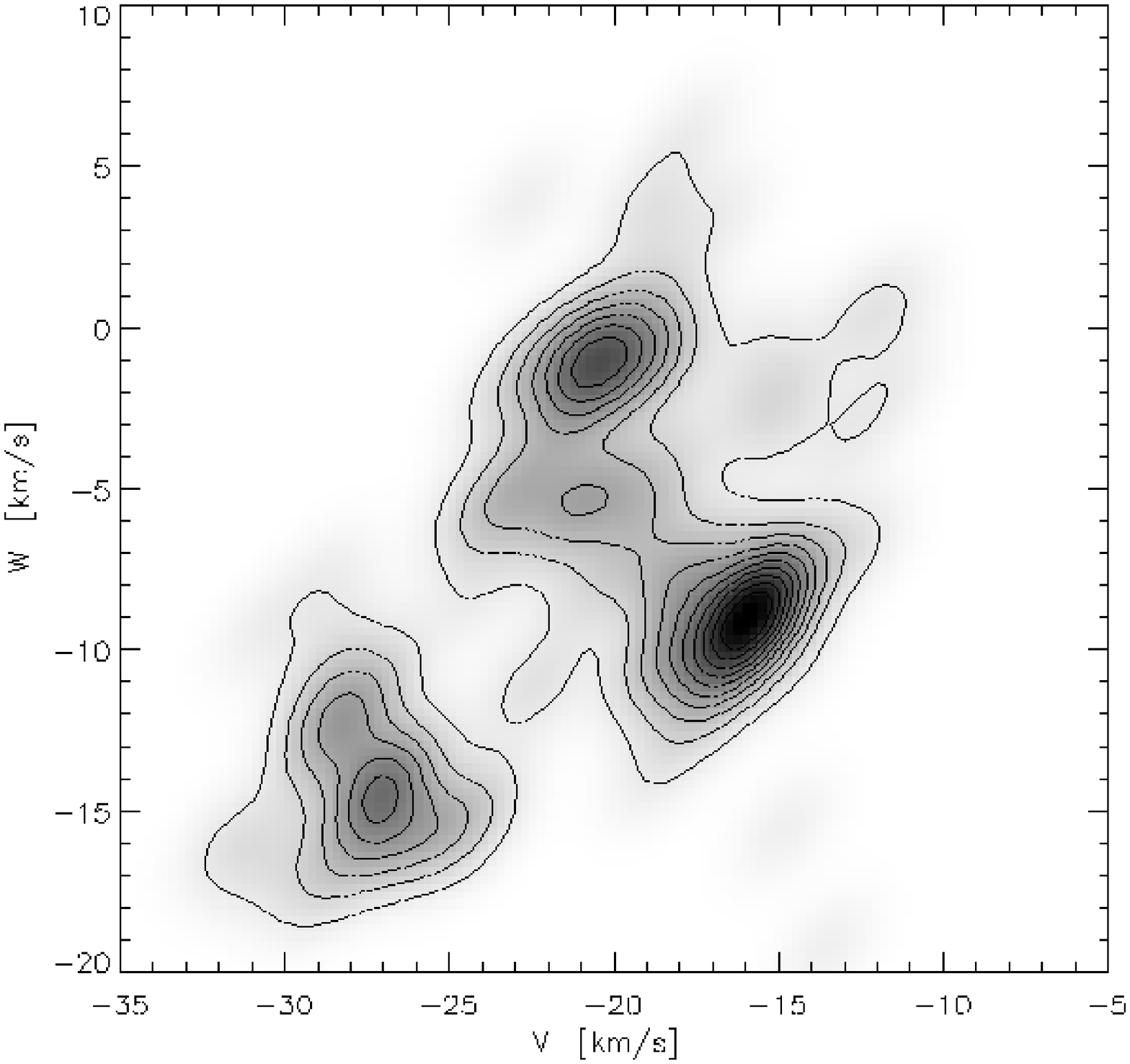}}
\hfill
\caption{Distribution in the ($U$-$V$) [left], ($U$-$W$) [middle] and 
         ($V$-$W$) [right] planes of heliocentric velocity components for 
         YLA member stars whose complete kinematic data are available. A 
         kernel estimator was used to indicate the isocontours.}
\label{fig.YLA.kernel}
\end{figure*}

The resulting YLA velocity components are very similar to those of the
associations in the Sco-Cen complex (see Table \ref{tab.Assoc}). Taking
into account the similar kinematics observed among YLA, and the observation 
that most of them are nearly coeval (ages of $\sim$5-15 Myr), it is interesting
to consider whether we can really speak of distinct entities. Figure
\ref{fig.YLA.kernel} shows the velocity distribution of YLA member stars
for which complete kinematic data are available. Using a kernel estimation,
several of the associations are clearly distinguished kinematically: we
can clearly identify peaks corresponding to Tuc-Hor/GAYA, $\beta$ Pic-Cap,
$\epsilon$ Cha and $\eta$ Cha. The HD 141569 system is not identified in Fig. 
\ref{fig.YLA.kernel} due to the small number of members whose complete 
kinematic data are available (only 2 stars).

Initially, it is surprising that the TW Hya association is not evident in
Fig. \ref{fig.YLA.kernel}, and even a local minimum in density can be 
observed near its expected positions ($[U,V,W] \sim [-10,-17,-5]$ km s$^{-1}$; 
see Table \ref{tab.Assoc}). This may be due to a combination of three factors: 
its position in the $(UVW)$ planes very near other YLA, its relatively high 
total velocity dispersion $\sigma$ (computed as $\sigma = \sqrt{ \sigma_U^2 + 
\sigma_V^2 + \sigma_W^2 }$ and compared with its neighbours in these planes), 
and the relatively low number of TW Hya stars with complete kinematic data 
($N_k$ in Table \ref{tab.Assoc}) compared to Tuc-Hor/GAYA and $\beta$ Pic-Cap 
associations (resulting in a low contribution to the probability density 
function shown in Fig. 3).

\begin{figure*}
\centering
\resizebox{6cm}{!}{\includegraphics{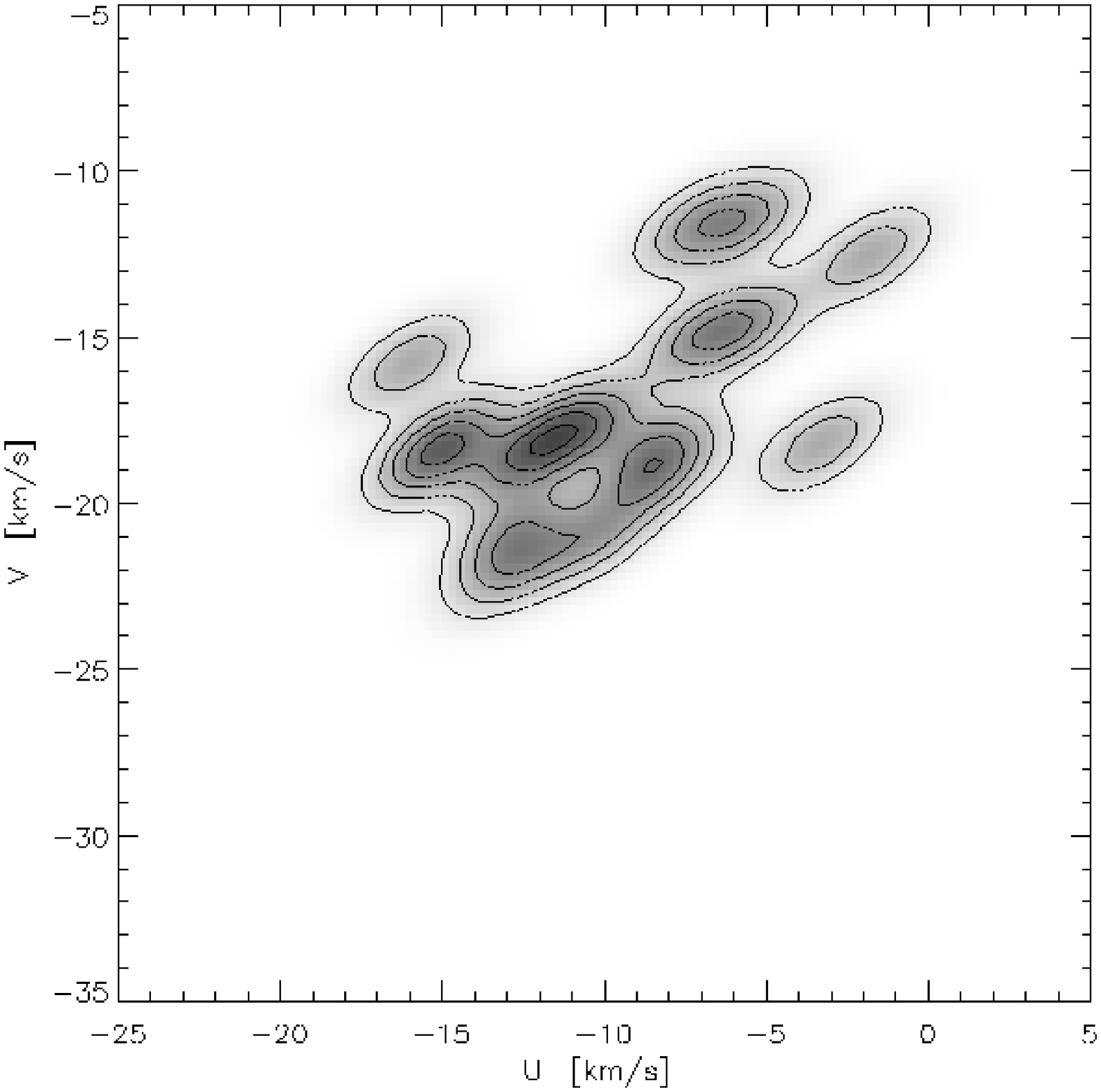}}
\hfill
\leavevmode
\resizebox{6cm}{!}{\includegraphics{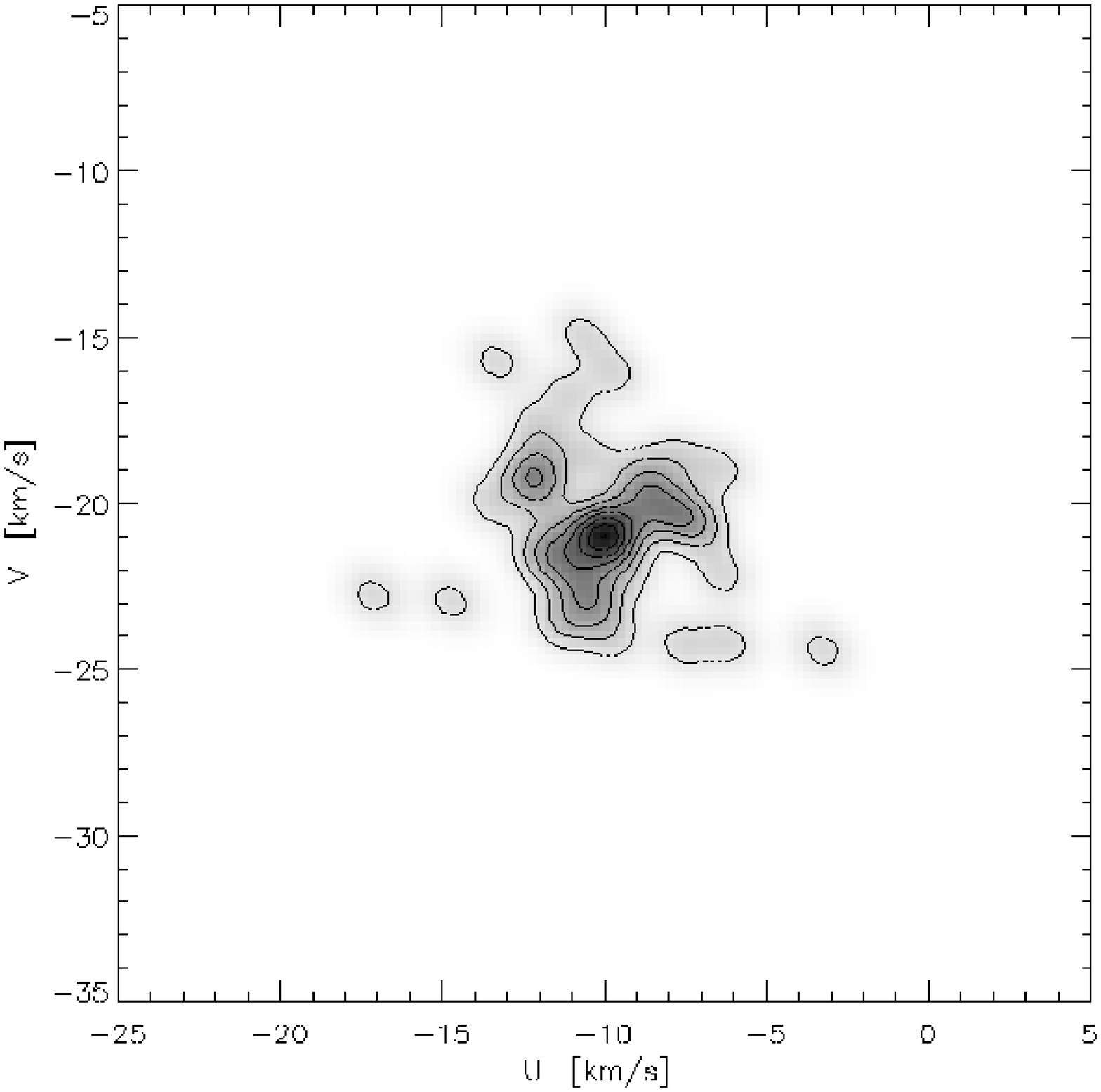}}
\hfill
\leavevmode
\resizebox{6cm}{!}{\includegraphics{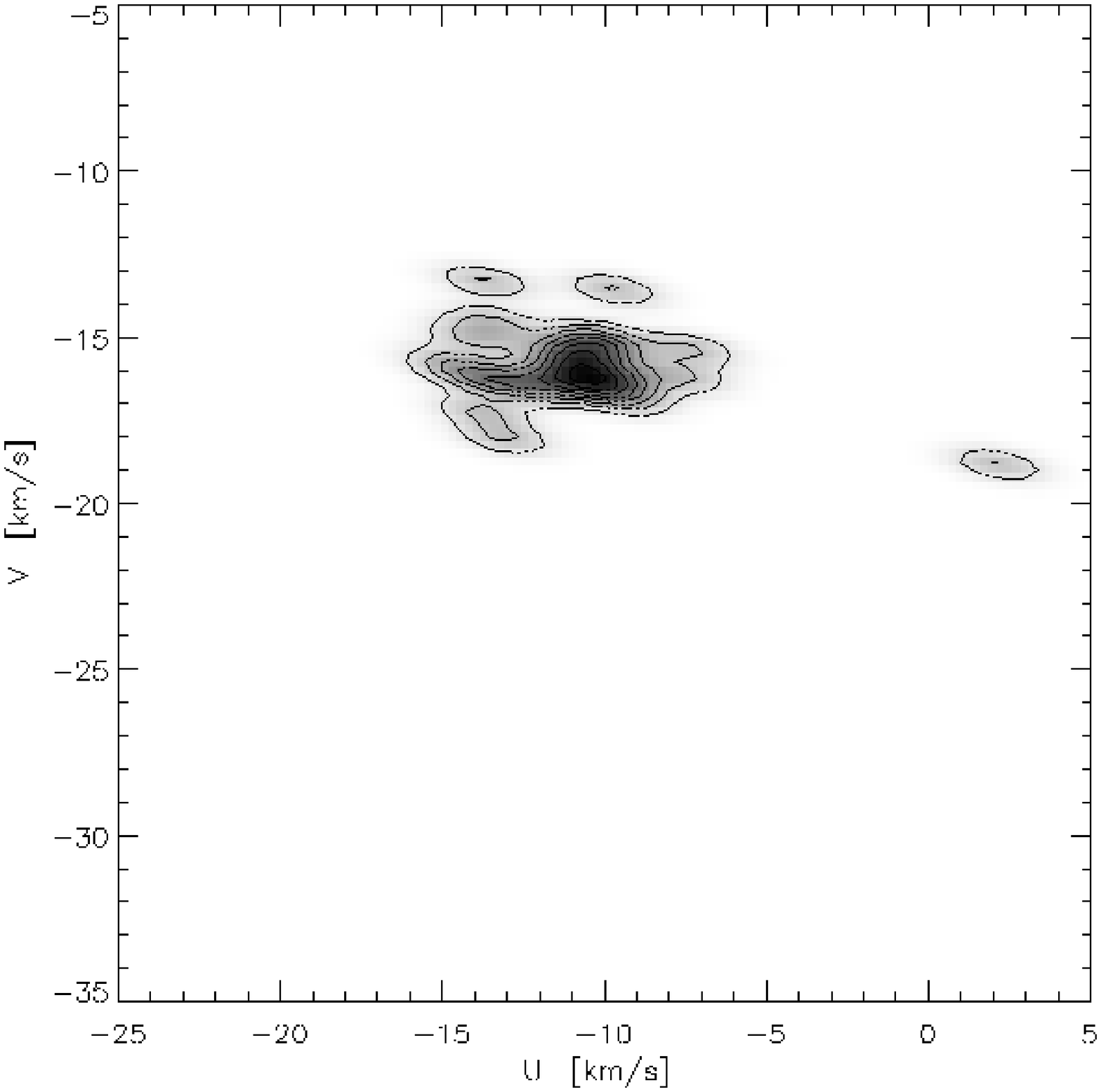}}
\vfill
\resizebox{6cm}{!}{\includegraphics{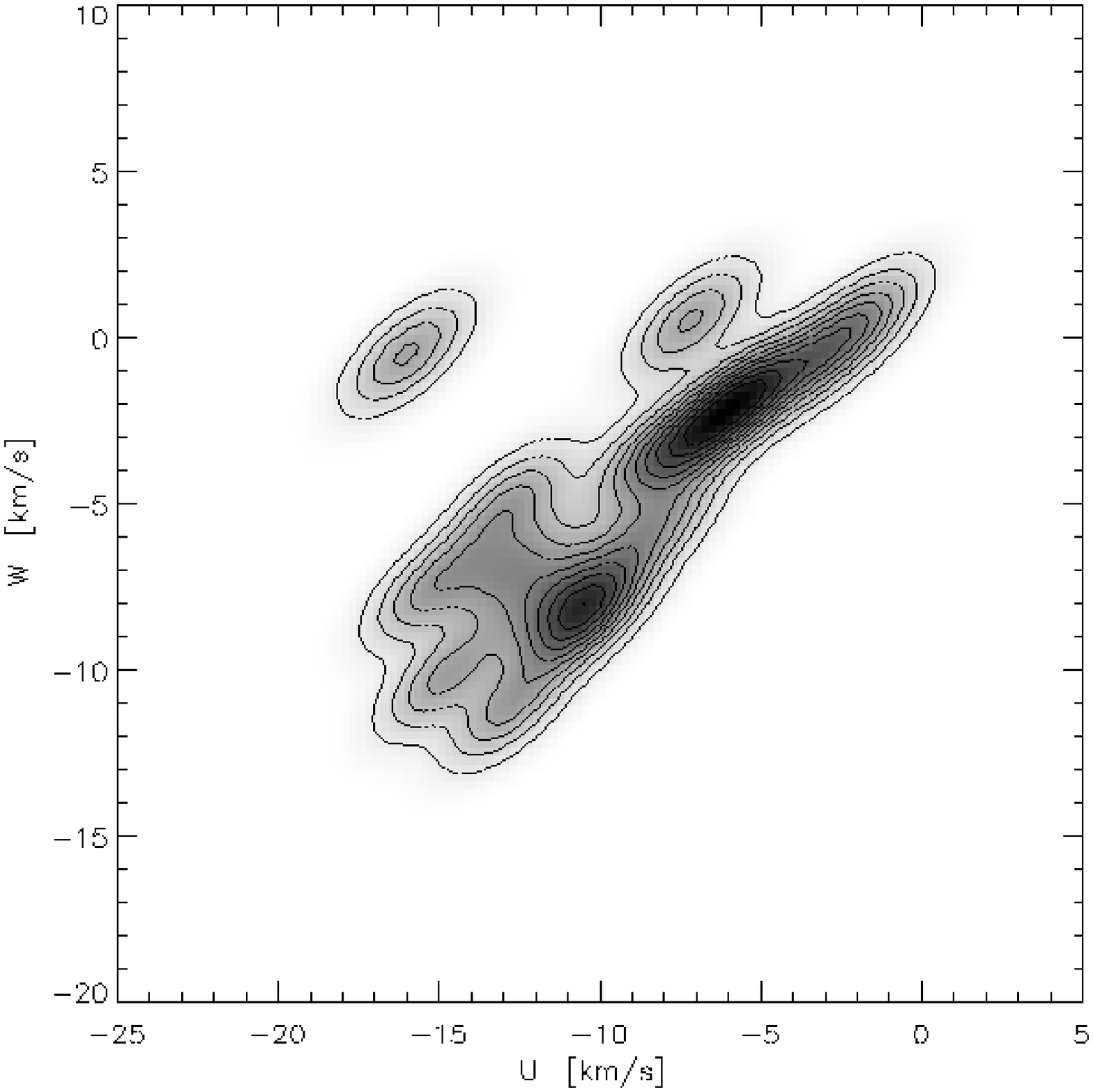}}
\hfill
\leavevmode
\resizebox{6cm}{!}{\includegraphics{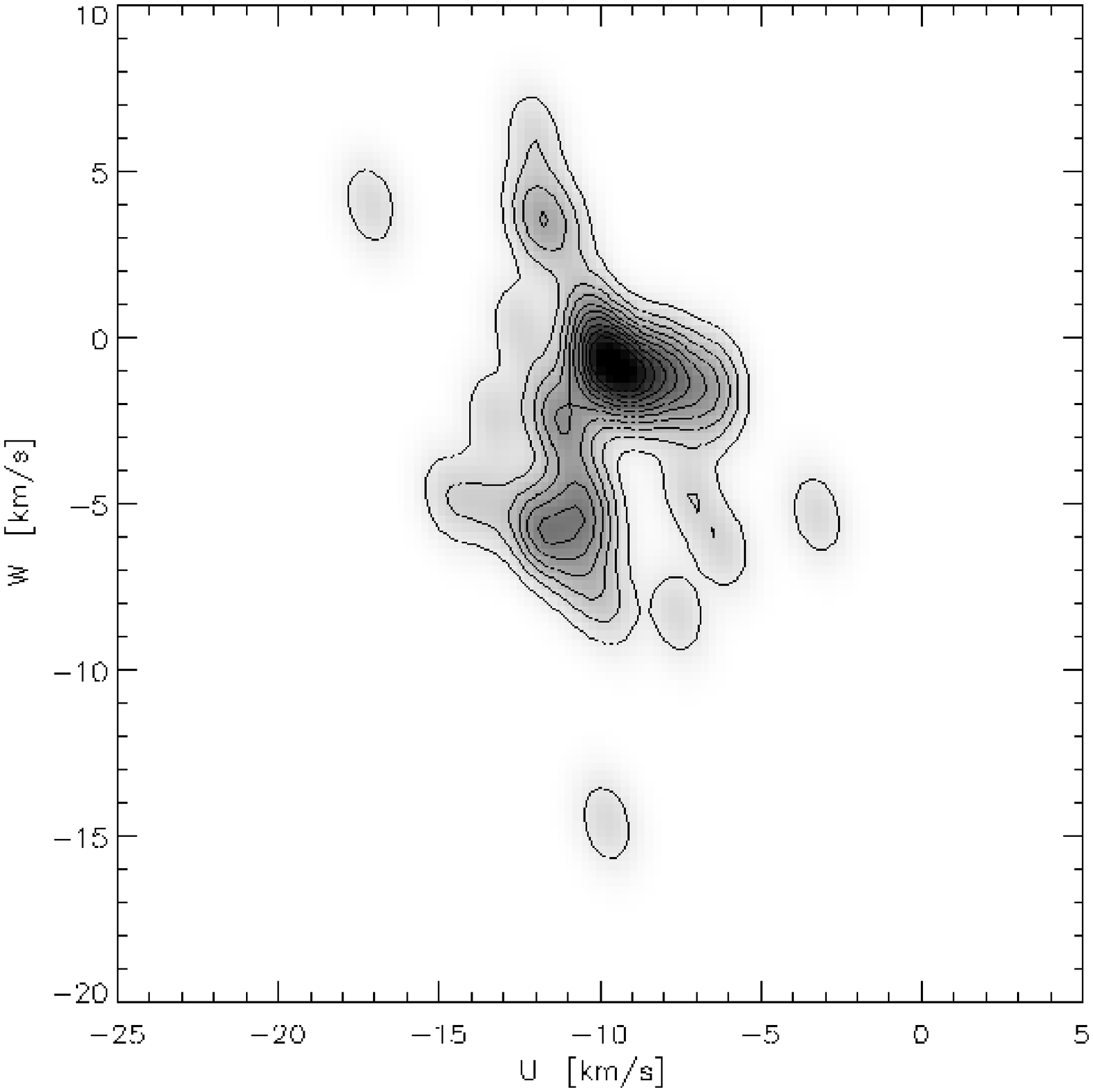}}
\hfill
\leavevmode
\resizebox{6cm}{!}{\includegraphics{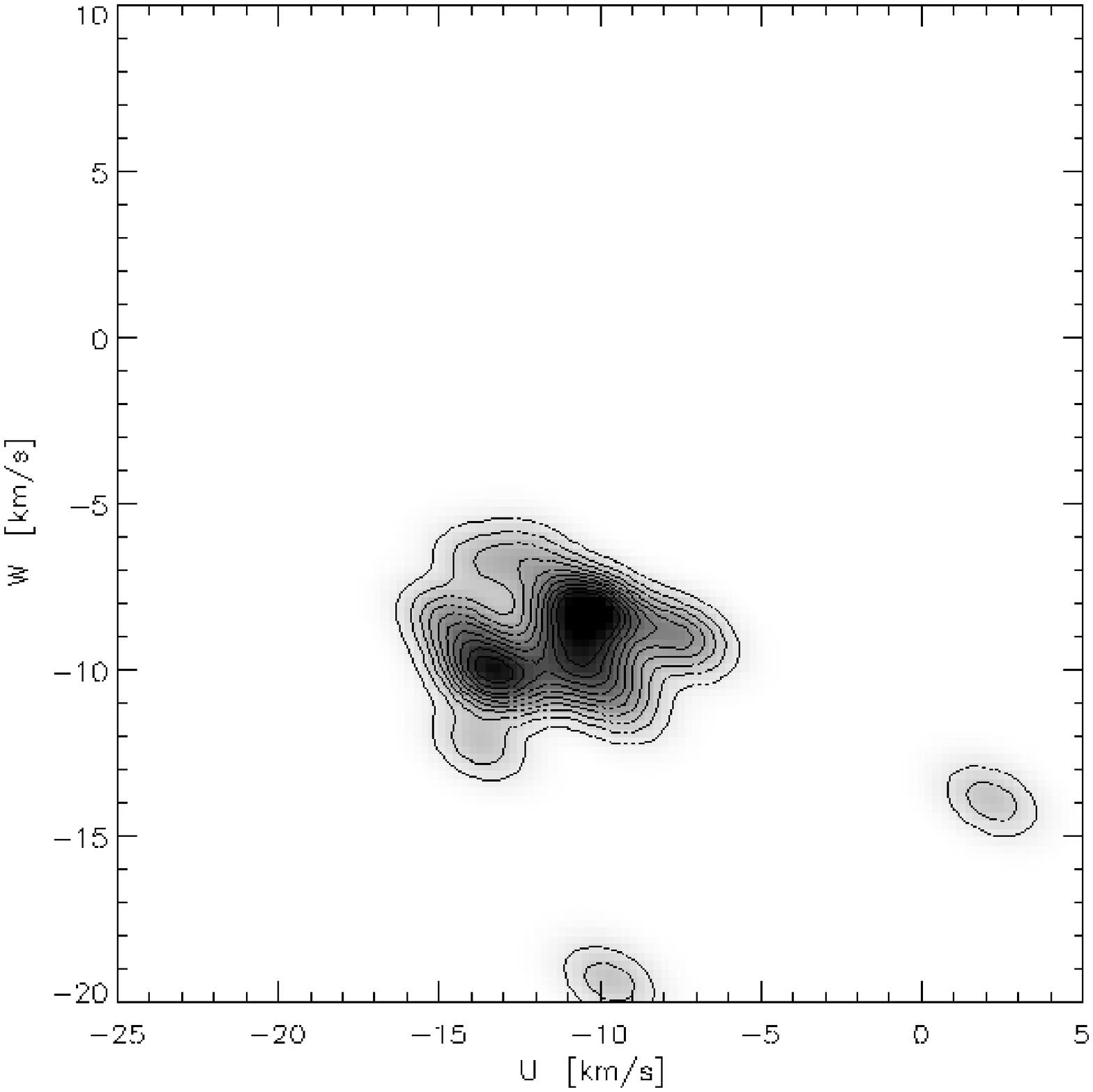}}
\vfill
\caption{Distribution in the ($U, V$) [top row] and ($U, W$) [bottom 
         row] planes of heliocentric velocity components for those stars 
         with complete kinematic data belonging to the TW Hya association 
         (left column), the Tuc-Hor/GAYA association (middle column) and 
         the $\beta$ Pic-Cap moving group (right column).}
\label{fig.YLA.sep.kernel}
\end{figure*}

\begin{figure}
\centering
\resizebox{8cm}{!}{\includegraphics{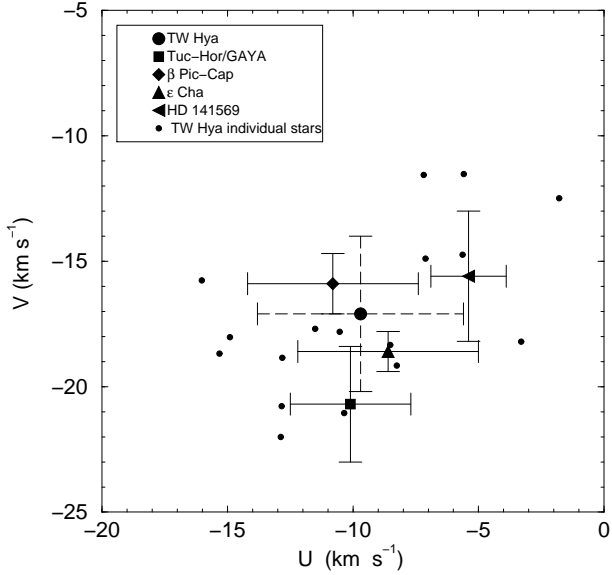}}
\caption{Mean helicocentric velocity components of the YLA (indicating 
         their 1$\sigma$ error bars; standard deviation) and position in the 
	 ($U, V$) plane of those stars (with complete kinematic data) belonging 
	 to the TW Hya association.}
\label{fig.TW_Hya_stars}
\end{figure}

To study this case in more detail, Fig. \ref{fig.YLA.sep.kernel} shows the
velocity distribution [$(U,V)$ and $(U,W)$ planes] of the stars belonging to 
the three associations with the most complete kinematic data: TW Hya, 
Tuc-Hor/GAYA and $\beta$ Pic-Cap. At first sight it is clear that the two
entities Tuc-Hor/GAYA and $\beta$ Pic-Cap are much better defined in these 
planes than the TW Hya association is. The main kernel estimator peaks agree 
with the mean values shown in Table \ref{tab.Assoc}, $(\overline{U},
\overline{V}, \overline{W})$ for these two associations. However, in the case 
of TW Hya, no central peak is found at the expected position on the planes for 
the mean values. In Fig. \ref{fig.TW_Hya_stars} we show the positions in the
($U,V$) plane for the stars belonging to the TW Hya association, and we compare 
them with the mean velocity components for the YLA. Whereas the other YLA have 
their own place in this plane (with low superposition of their error bars), the 
TW Hya association seems to fill a region shared by other YLA. We have computed 
the distance in the $(UVW)$ velocity space from each TW Hya star to the 
position of the mean velocity of the YLA. Only 4 out of 17 stars are nearer to 
the position of the TW Hya association in the $(UVW)$ space than to other YLA. 
In Table \ref{tab.chg_vel_YLA} we have done the exercise of recomputing the 
mean velocity components and their dispersions for the YLA after adding the TW 
Hya stars to the nearest YLA in the $(UVW)$ space. We can see that the change 
in the mean velocity vectors of the associations is much smaller than their 
corresponding standard deviation. A controversial hypothesis one could propose 
is that the TW Hya association (the first one discovered among the YLA) could 
not actually be an association at all, and one may even wonder if its member 
stars could be redistributed among the other YLA. However, there are two 
important facts that do not support this hypothesis. The first one concerns 
the spatial distribution of the YLA at present. As can be seen in Fig. 
\ref{fig.YLA}, TW Hya is clearly isolated in the $(X,Z)$ plane (not considering 
the much older members of the AB Dor moving group), with a spatial extent in 
agreement with that observed in the other YLA. The previously discussed 
kinematic reassigment of stars would enlarge, clearly in excess, the present 
accepted size for the other associations. The second important fact is that, 
whereas almost all Tuc-Hor/GAYA and $\beta$ Pic-Cap members have Hipparcos 
parallaxes, 70\% of the TW Hya members have only an estimated distance from 
broadband photometry (Song et al. \cite{Song03}). The assumption of a relative 
error of about 20\% for this parameter would have a contribution of up to 3-4 
km s$^{-1}$ in the velocity dispersion components. This value, convolved with 
an intrisic velocity dispersion of about 1 km s$^{-1}$ (Mamajek 
\cite{Mamajek05}), would justify the values shown in Table 1. Furthermore, we 
have confirmed that the discrepancies between these photometric distances and 
the moving cluster distances by Mamajek (\cite{Mamajek05}) are about 20\%. From 
the above arguments it is clear that more accurate astrometric data is 
mandatory to clarify both the kinematic dispersion and membership assignment of 
TW Hya members. The large dispersion observed in the estimated ages for this 
association (see Table \ref{tab.AssocAges}) also contributes to the deficient 
characterisation of this entity.

\begin{table}
   \caption{Mean heliocentric velocity components (and their standard
            deviations) of the young local associations before and after adding
	    those stars belonging to TW Hya (see text).}
   \label{tab.chg_vel_YLA}
\centering
\begin{tabular}{lrrrr}
\hline
\hline
Association     & $\overline{U}$ & $\overline{V}$ & $\overline{W}$
                & $N_\mathrm{k}$ \\
                & (km s$^{-1}$) & (km s$^{-1}$) & (km s$^{-1}$) & \\
\hline
Tuc-Hor/GAYA    & $-10.1_{(2.4)}$ & $-20.7_{(2.3)}$ &  $-2.5_{(3.8)}$
                & 44 \\
adding stars    & $-10.2_{(2.6)}$ & $-20.5_{(2.3)}$ &  $-2.5_{(3.7)}$
                & 47 \\
\hline
$\beta$ Pic-Cap & $-10.8_{(3.4)}$ & $-15.9_{(1.2)}$ &  $-9.8_{(2.5)}$
                & 24 \\
adding stars    & $-11.2_{(3.3)}$ & $-16.3_{(1.4)}$ &  $-9.2_{(1.6)}$
                & 28 \\
\hline
$\epsilon$ Cha  &  $-8.6_{(3.6)}$ & $-18.6_{(0.8)}$ &  $-9.3_{(1.7)}$
                & 5 \\
adding stars	&  $-9.5_{(3.0)}$ & $-19.2_{(1.4)}$ &  $-8.9_{(1.9)}$
                & 9 \\
\hline
HD 141569       &  $-5.4_{(1.5)}$ & $-15.6_{(2.6)}$ &  $-4.4_{(0.8)}$
                & 2 \\
adding stars	&  $-5.2_{(1.9)}$ & $-14.3_{(2.5)}$ &  $-1.9_{(1.9)}$
                & 8 \\
\hline
\end{tabular}
\end{table}

%

\section{Stellar orbits} 
\label{sect.orbits}

The integration back in time of the Sco-Cen and YLA orbits allows us to
study their origin and possible influence on the immediate ISM over the
last million years. To compute the stellar orbits back in time we used the
code developed by Asiain et al. (\cite{Asiain99b}) based on the
integration of the equations of motion using a realistic model of the
Galactic gravitational potential.

If we consider a coordinate system ($\xi,\eta,\zeta$)\footnote{$\xi$
points towards Galactic centre, $\eta$ in the direction of Galactic
rotation and $\zeta$ towards the Galactic north pole.} centred on the Sun
and rotating around Galactic centre with a constant angular velocity
$\Omega_{\sun}$, the equations of motion of a star are:
\begin{eqnarray} \label{eq.motion}
  {\ddot \xi}   & = & - \displaystyle \frac{\partial \Phi}{\partial \xi} -
                        \Omega_{\sun}^2 (R_{\sun} - \xi) - 2 
                        \Omega_{\sun} \dot{\eta}
  \nonumber \\
  {\ddot \eta}  & = & - \displaystyle \frac{\partial \Phi}{\partial \eta} +
                        \Omega_{\sun}^2 \eta + 2 \Omega_{\sun} \dot{\xi}
  \nonumber \\
  {\ddot \zeta} & = & - \displaystyle \frac{\partial \Phi}{\partial \zeta}
\end{eqnarray}

\noindent where $\Phi = \Phi(R,\theta,z;t)$ is the gravitational potential 
of the Galaxy in galactocentric cylindrical coordinates.

When $\Phi$ is a known function, these equations can be solved numerically
through a fourth-order Runge-Kutta integrator. We have decomposed $\Phi$
into three components: the general axisymmetric potential
$\Phi_\mathrm{AS}$ (Allen \& Santill\'an \cite{Allen91}), the potential
due to the spiral structure of the Galaxy $\Phi_\mathrm{Sp}$, and that due
to the central bar $\Phi_\mathrm{B}$ (see Table \ref{tab.parorb}).  In
this way we get a realistic estimate of the Galactic gravitational
potential.

The spiral structure potential is taken from Lin's theory (Lin \& Shu
\cite{Lin64}; Lin et al. \cite{Lin69}):

\begin{equation}\label{eq.mot}
  \Phi_\mathrm{Sp} (\mbox{R},\theta;t) =
  {\cal A} \cos \left[ m (\Omega_p t - \theta) + \psi(R) \right]
\end{equation}

\noindent where:
\begin{eqnarray}
  {\cal A}  & = & \displaystyle \frac {(R_{\sun} \Omega_{\sun})^2 
                  f_\mathrm{r0} \tan i}{m}
\nonumber \\
  \psi(R) & = & - \displaystyle \frac {m}{\tan i}
                    \ln \left( \displaystyle \frac {R}{R_{\sun}} \right) +
                    \psi_{\sun}
\end{eqnarray}

\noindent where ${\cal A}$ is the amplitude of the spiral potential,
$f_\mathrm{r0}$ the ratio between the radial component of the force due to
the spiral arms and that due to the general Galactic field, $\Omega_p$ the
constant angular velocity of the spiral pattern, $m$ the number of spiral
arms, $i$ the pitch angle of the arms, $\psi$ the radial phase of the
spiral wave and $\psi_{\sun}$ its value at the position of the Sun. We
have taken the values obtained by Fern\'andez et al. (\cite{Fernandez01};
see also Fern\'andez \cite{Fernandez05}) for a model with 2 spiral arms.

In the case of the potential due to the central bar, we have chosen the
triaxial ellipsoid model of Palou\u s et al. (\cite{Palous93}) with a
rotation speed $\Omega_\mathrm{B} = 70$ km s$^{-1}$ kpc$^{-1}$ (Binney et
al. \cite{Binney91}). Although large uncertainties are still present in
the bar parameters, this structure has an almost negligible effect on our
stellar trajectories (as far back as 30 Myr; see Asiain \cite{Asiain98}).

\begin{table}
\caption{Position and velocity of LSR, and local density (Kerr \& 
         Lynden-Bell \cite{Kerr86}), bulge, disk and halo parameters 
         (Miyamoto \& Nagai \cite{Miyamoto75}; Allen \& Santill\'an 
         \cite{Allen91}), together with spiral structure (Fern\'andez et al.
         \cite{Fernandez01}) and central bar (Binney et al. 
         \cite{Binney91}; Palou\u s et al. \cite{Palous93}) parameters 
         used in the integration of stellar orbits.}
\label{tab.parorb}
\centering
\begin{tabular}{ll}
        \hline
$R_{\sun}$            &  8.5 kpc \\
$\Theta_{\sun}$       & 220 km s$^{-1}$ \\
$\rho_{\sun}$         &  0.15  M$_{\sun}$ \, pc$^{-3}$   \\
        \hline
$M_\mathrm{B}$        &  $1.4\cdot10^{10}$ M$_{\sun}$ \\
$M_\mathrm{D}$        &  $8.6\cdot10^{10}$ M$_{\sun}$ \\
$M_\mathrm{H}$        &  $1.1\cdot10^{11}$ M$_{\sun}$ \\
$a_\mathrm{B}$        &  0.39 kpc \\
$a_\mathrm{D}$        &  5.3  kpc \\
$b_\mathrm{D}$        &  0.25 kpc \\
$a_\mathrm{H}$        & 12.0  kpc \\
        \hline
$m$                   &     2       \\
$i$                   &   $-6\degr$ \\
$\psi_{\sun}$         &  330$\degr$ \\
$f_{\mathrm{r}}$      &    0.05     \\
$\Omega_{\mathrm{p}}$ & 30 km s$^{-1}$ kpc$^{-1}$ \\
\hline
${a_\mathrm{B}}/{b_\mathrm{B}}$ \hspace{0.5cm} &    2.381    \\
\vspace{1mm}
${a_\mathrm{B}}/{c_\mathrm{B}}$ &    3.030    \\
\vspace{1mm}
$q_\mathrm{B}$        &     5 kpc   \\
$t_o$                 &    $5 \cdot 10^8$ years \\
$\theta_o$            & 45\degr     \\
$\Omega_\mathrm{B}$   & 70 km s$^{-1}$ kpc$^{-1}$ \\
\hline
\end{tabular}
\end{table}

\begin{figure}
\centering
\resizebox{7cm}{!}{\includegraphics{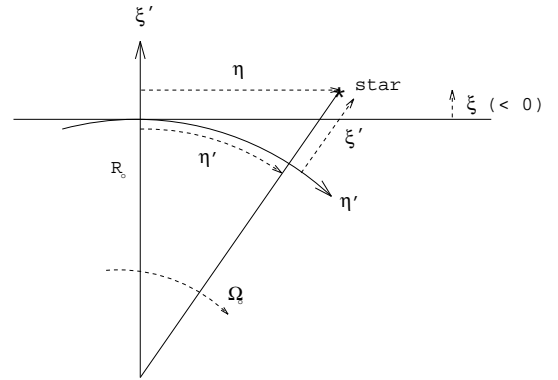}}
\caption{Heliocentric coordinates ($\xi,\eta,\zeta$) and 
         ($\xi^\prime,\eta^\prime,\zeta^\prime$) (from Asiain 
         \cite{Asiain98}; see also Asiain et al. \cite{Asiain99b}).}
\label{fig.Ricard_Coor2}
\end{figure}

All the figures in the next section use the ($\xi^\prime, \eta^\prime,
\zeta^\prime$) coordinate system centred on the Sun's current position and
rotating around Galactic centre at a constant velocity $\Omega_{\sun}$
(see Fig. \ref{fig.Ricard_Coor2}). $\xi^\prime$ points to the
Galactic anti-centre and is defined as $\xi^\prime = R - R_{\sun}$.  
$\eta^\prime$ is a linear coordinate, measured along the circle of radius
$R_{\sun}$, which is positive in the direction of Galactic rotation. The
coordinate system ($\xi^\prime, \eta^\prime, \zeta^\prime$) is convenient
for our purposes as it minimises variation in the variables.

\begin{figure*}
\centering
\resizebox{16cm}{!}{\includegraphics{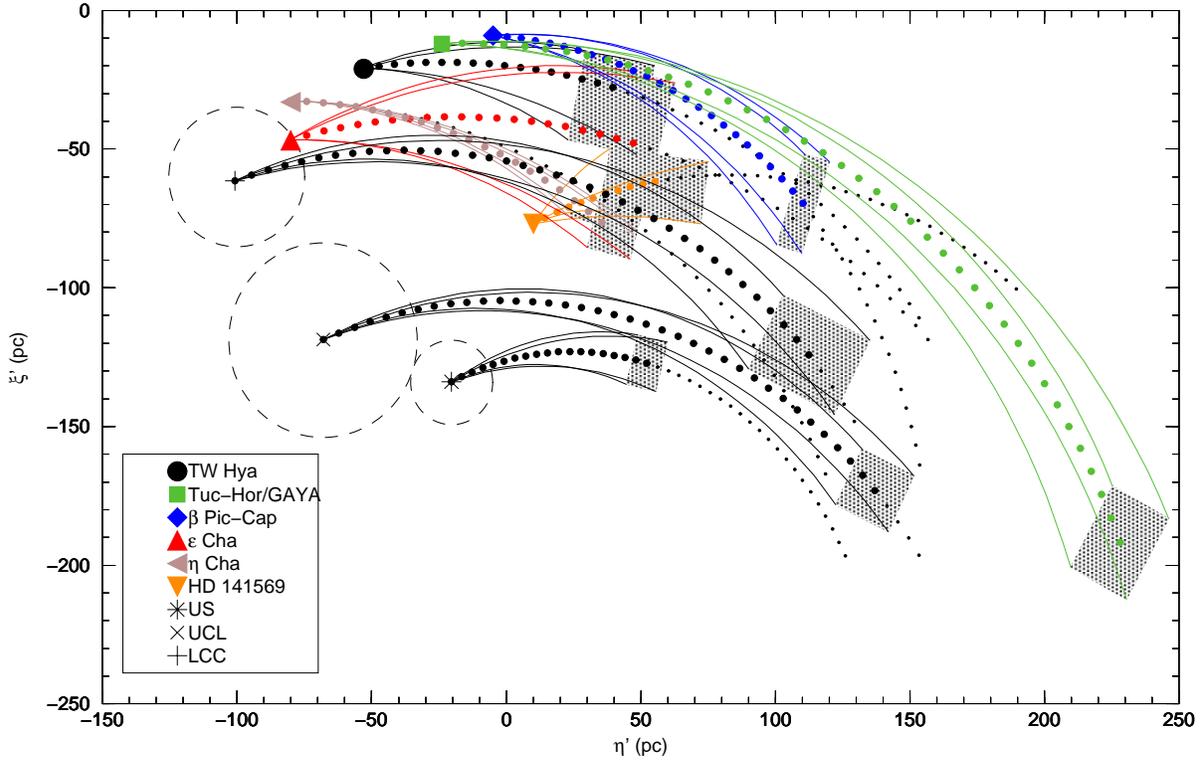}}
\caption{Positions and orbits in the Galactic plane ($\xi^\prime,\eta^\prime$) 
         of YLA and the Sco-Cen complex going back in time to their individual 
	 ages. The orbits are also shown (small dots) as far back as $-20$ Myr.
	 The grey areas show the expected positional errors at birth due
	 to kinematic observational errors. The centre of the ($\xi^\prime,
	 \eta^\prime$) coordinate system is comoving with the LSR.}
\label{fig.Associacions+ScoCen3_xi-eta}
\end{figure*}

\begin{figure}
\centering
\resizebox{\hsize}{!}{\includegraphics{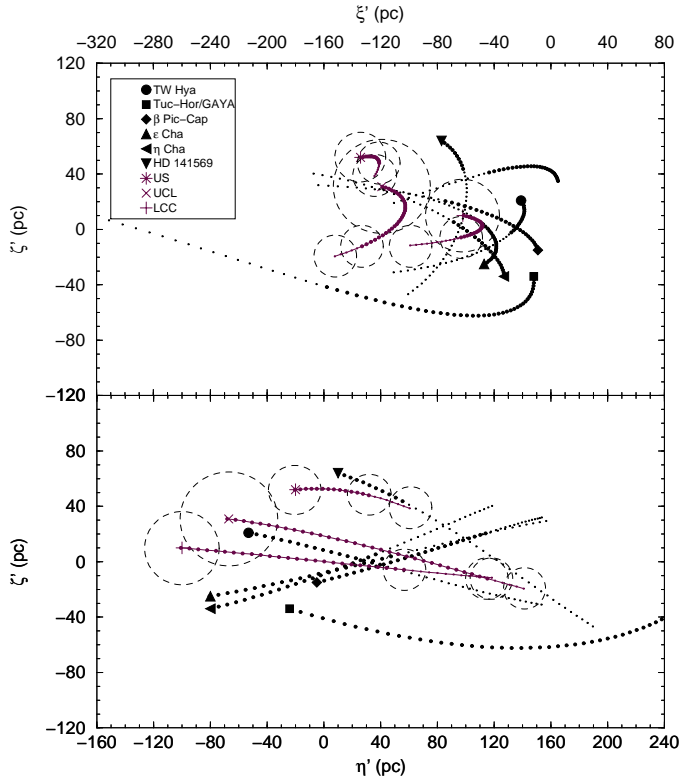}}
\caption{Present positions and orbits in the Galactic planes
         ($\xi^\prime,\zeta^\prime$) (top) and 
         ($\eta^\prime,\zeta^\prime$) (bottom) for YLA and the Sco-Cen complex. 
          The {\it error areas} (not shown) are of similar size as in Fig. 
          \ref{fig.Associacions+ScoCen3_xi-eta}. (See comments at Fig. 
         \ref{fig.Associacions+ScoCen3_xi-eta}.)}
\label{fig.Associacions+ScoCen3_xi-eta-zeta}
\end{figure}

We have integrated the orbits of the associations as a whole, using the mean
position and velocity for each association shown in Table \ref{tab.Assoc}. The
results are presented in Figs. \ref{fig.Associacions+ScoCen3_xi-eta} and
\ref{fig.Associacions+ScoCen3_xi-eta-zeta}, where not only do we consider the
orbits in the Galactic plane (as is usual in the literature), but also in the
meridian and rotation planes. In Fig. 
\ref{fig.Associacions+ScoCen3_xi-eta} we show the estimated error in the 
position of the associations at birth (grey area). These {\it error areas} have
been obtained computing the orbits back in time, using the values $\overline{U} \pm 2 Se_{\overline{U}}$, $\overline{V} \pm
2 Se_{\overline{V}}$ as present velocity components (where $Se_{\overline{U}}$ 
and $Se_{\overline{V}}$  are the standard errors in the $\overline{U}, 
\overline{V}$ mean velocity  components obtained from the standard deviations 
in Table \ref{tab.Assoc}). We have confirmed that uncertainties in the 
$\overline{W}$ component have a negligible effect on the orbits shown in Fig. 
\ref{fig.Associacions+ScoCen3_xi-eta} \footnote{ A variation of $\pm 2 
Se_{\overline{W}}$ in $\overline{W}$ implies a change in the vertical position 
($\zeta^\prime$) with time, thus slightly affecting the radial and azimuthal 
forces acting on the star (see Eq. 9 in Asiain et al. \cite{Asiain99b} and Eqs. 
1, 3 and 5 in Allen \& Santill\'an \cite{Allen91}). However, we have confirmed 
that this effect results in a change in position of less than $10^{-3}$ pc in 
the $(\xi^\prime,\eta^\prime)$ plane in the present case.}. An error in age 
shall be read as a displacement of the error areas along the plotted orbits in 
the figure. Unfortunately, age uncertainties are large for all the associations 
(see Table \ref{tab.AssocAges}). Later in this paper we study how these large 
uncertainties in age may affect our results.

The most obvious trend observed in Figs. \ref{fig.Associacions+ScoCen3_xi-eta} 
and \ref{fig.Associacions+ScoCen3_xi-eta-zeta} is the spatial concentration of 
all the associations (Sco-Cen complex and YLA) in the first Galactic quadrant 
in the past. All the YLA (except the relatively {\it old} Tuc-Hor/GAYA) were in 
the region of the Galactic plane bounded by $-70 \la \xi^\prime \la -30$ pc and 
$35 \la \eta^\prime \la 110$ pc at their time of birth. There is a very 
conspicuous spatial grouping at the time of birth of TW Hya, $\epsilon$ Cha, 
$\eta$ Cha and HD 141569 (in a sphere 25 pc in radius). At its birth, $\beta$ 
Pic-Cap was located about 50 pc from the other YLA. All the associations were 
concentrated in the region 0 $\la \zeta^\prime \la$ 45 pc at their birth.
Therefore the associations were formed slightly above the Galactic plane (the 
Sun, whose position is the origin of the $\zeta^\prime$ coordinate, is located 
about 16 pc above the Galactic plane; see Binney \& Merrifield 
\cite{Binney98}). The region where these associations were formed had a size of 
40 x 75 x 45 pc ($\xi^\prime$ x $\eta^\prime$ x $\zeta^\prime$). At present, 
the stars belonging to YLA are distributed in a region of about 120 x 130 x 140 
pc. The volume they occupy has therefore increased by a factor of $\sim16$
since their birth. As can be observed in Fig. 
\ref{fig.Associacions+ScoCen3_xi-eta}, the errors associated with the mean 
velocity components of the associations do not have a crucial influence 
on the previous results: the error areas have typical side lengths of about 
10-30 pc.

In the case of the Sco-Cen associations, both UCL and LCC began life very
near the Galactic plane; for both of them $-19 \la \zeta^\prime \la -12$
pc at birth (they were closer than $\sim$4 pc to the Galactic plane). For
US, $\zeta^\prime \sim$40 pc at birth.

%

\section{Origin and evolution of the local structures}
\label{sect.origin}

The results obtained in the previous section suggest that the formation
of YLA was triggered in a neighbouring region of the first
Galactic quadrant between 5 and 15 Myr ago. In this section we present a
description of this mechanism and link it to the origin and evolution
of the Sco-Cen complex and the LB.

\subsection{Origin of the Sco-Cen complex}\label{sect.origin.sco-cen}

The gas associated with the Sco-Cen complex has classically been
considered as a part of the Lindblad Ring; a toroidal structure of HI and
molecular clouds in expansion (Lindblad \cite{Lindblad67}) belonging to
the Gould Belt. Close to Sco-Cen, the clouds of Lupus and $\rho$
Oph, and the so-called Aquila Rift, would also form part of the Lindblad
Ring.

The structure that results from considering all these molecular complexes
has a length of about 120\degr\,, and extends from the Vela region to the
Aquila Rift. Below, we discuss the most plausible scenarios for star
formation in this region and we conclude that the most likely is the
collision of a GMC with a spiral arm. We mention the review published by
S03, where other less plausible scenarios can be found.

\subsubsection{The sequential star-formation model}

This model was formulated by Blaauw (\cite{Blaauw64}, \cite{Blaauw91}) and
suggests that star formation begins at one end of a GMC and propagates to
adjacent regions due to stellar winds and supernovae.

Preibisch \& Zinnecker (\cite{Preibisch99}) later proposed a history of
star formation in the Sco-Cen complex, using the {\it classical} ages of
de Geus et al. (\cite{Geus89}). Star formation in Sco-Cen would have begun
around 15 Myr ago in UCL. 12 Myr ago, when star formation started in LCC,
the most massive star in UCL would have exploded as a supernova, forming
the largest HI shell surrounding the association. 5 Myr ago, the shock
front from this supernova would have passed through the cloud, which was to
become the parent of US, triggering star formation there. Shortly after,
the strong stellar winds from the most massive stars in US would begin to
sweep away the molecular cloud, stopping the star formation process. Only
1.5 Myr ago, the most massive star in US would have exploded as a
supernova, completely dispersing the US cloud. Nowadays, the wave front of
this supernova would be travelling through the $\rho$ Oph molecular cloud,
triggering star formation there.

Following S03, this model would result in systematic velocity fields with 
radial components directed away from the different centres of star formation. 
This is not what we observe in the results obtained in previous sections. 
However, the problem could be a little more complicated since the velocities of
the young stars would not only depend on the velocity and direction of the 
shock front that compressed the parent cloud; the kinematics of the new stars 
would also depend on the original velocity of the molecular cloud, which could 
be similar to, or even larger than, the velocity of the shock front. Following 
this, the successive sources of star formation would not necessarily form 
groups of stars with clearly identified radial velocity components. Thus, 
contrary to the opinion of S03, we think that the sequential star-formation 
mechanism could have a crucial role in the global scenario for star formation 
in this region.

\subsubsection{The Gould Belt model based on an expanding gas ring}

The kinematics of the stars belonging to the Gould Belt seems compatible with 
the expansion model proposed by Olano (\cite{Olano82}): a gas ring with a 
centre located at a distance of about 166 pc in the direction $l\sim131\degr$. 
However, the Sco-Cen complex is precisely one of the regions of the Gould Belt 
that does not fit this model, since the observed proper motions are not 
orientated as predicted by Olano (see also Moreno et al. \cite{Moreno99}). 
Moreover, as shown above, as we project the orbits of the stars in this region 
back in time, they no longer indicate the centre of the Gould Belt. S03 also 
mentions other problems, such as the fact that the age of the stars in Sco-Cen 
is an important fraction of the age of the Gould Belt predicted by this model 
(30 Myr), in spite of the large distance between the present position of 
Sco-Cen and the expansion centre of the Belt.

Furthermore, as shown in a previous paper (Torra et al. \cite{Torra00}), the 
Sco-Cen associations are not needed to explain the peculiar kinematics of the 
Gould Belt.

\subsubsection{Star formation triggered by a spiral arm shock wave}

The region in question could also have been formed as a result of the
interaction of the parent GMC with the shock wave of a spiral arm. In this
case, the main problem to be faced is that the classical view of the
spiral structure of the Milky Way places the Sun in an interarm region, at
about 1 kpc from the nearest arm (see, for example, Georgelin \& Georgelin
\cite{Georgelin76}) where there is no spiral shock wave capable of
triggering star formation. However, in recent years there has been some
evidence that this classical view could be wrong (see, for instance,
Fern\'andez et al. \cite{Fernandez01} for a stellar kinematics study or
Am\^ores \& L\'epine \cite{Amores05} for a model of interstellar
extinction in the Galaxy), with the Sun closer to the inner
(Sagittarius-Carina) arm. This forces us to study this possible scenario
in more detail.

\begin{figure}
\resizebox{\hsize}{!}{\includegraphics{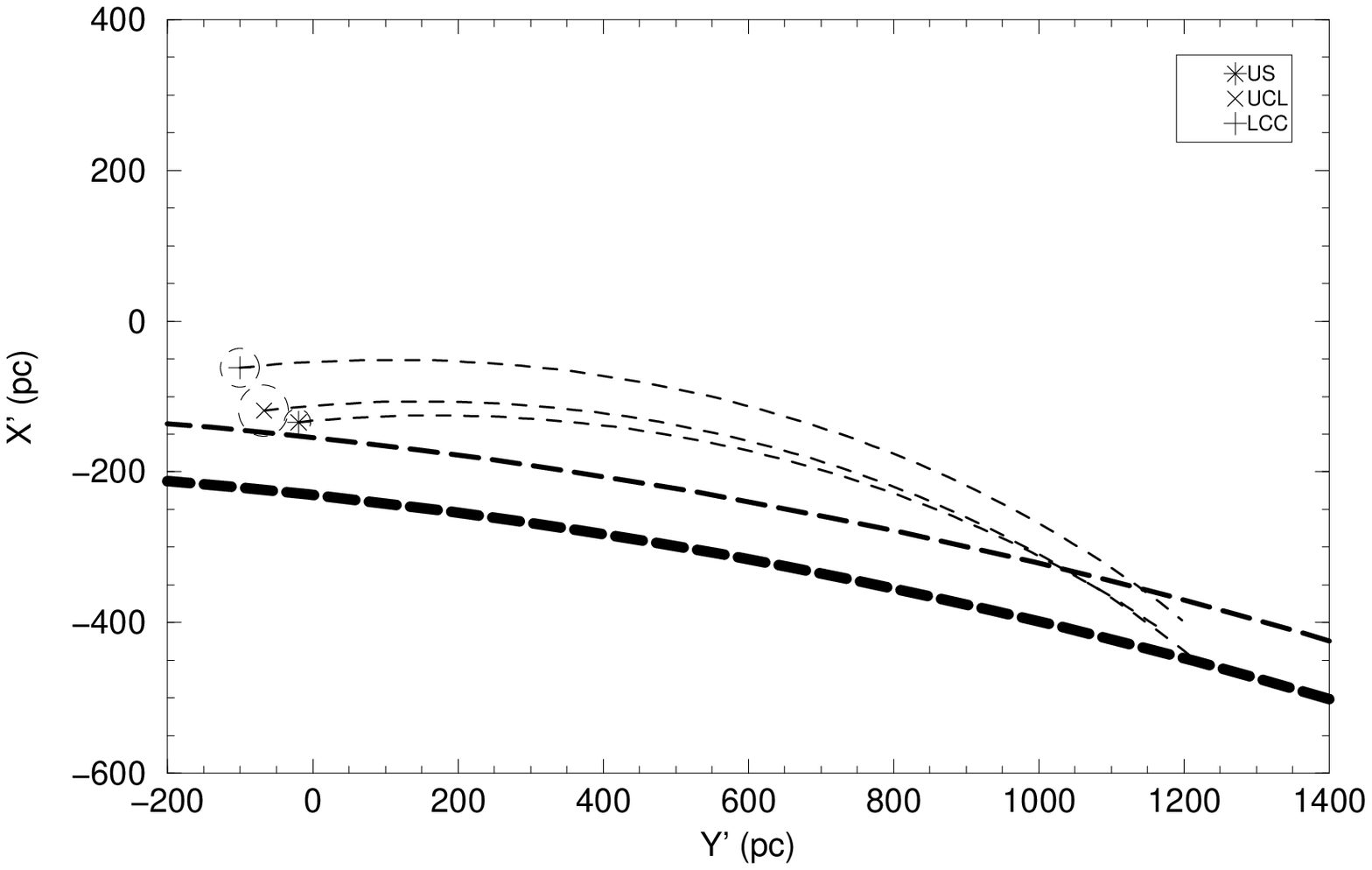}}
\hfill
\leavevmode
\vspace{0.5cm}
\resizebox{\hsize}{!}{\includegraphics{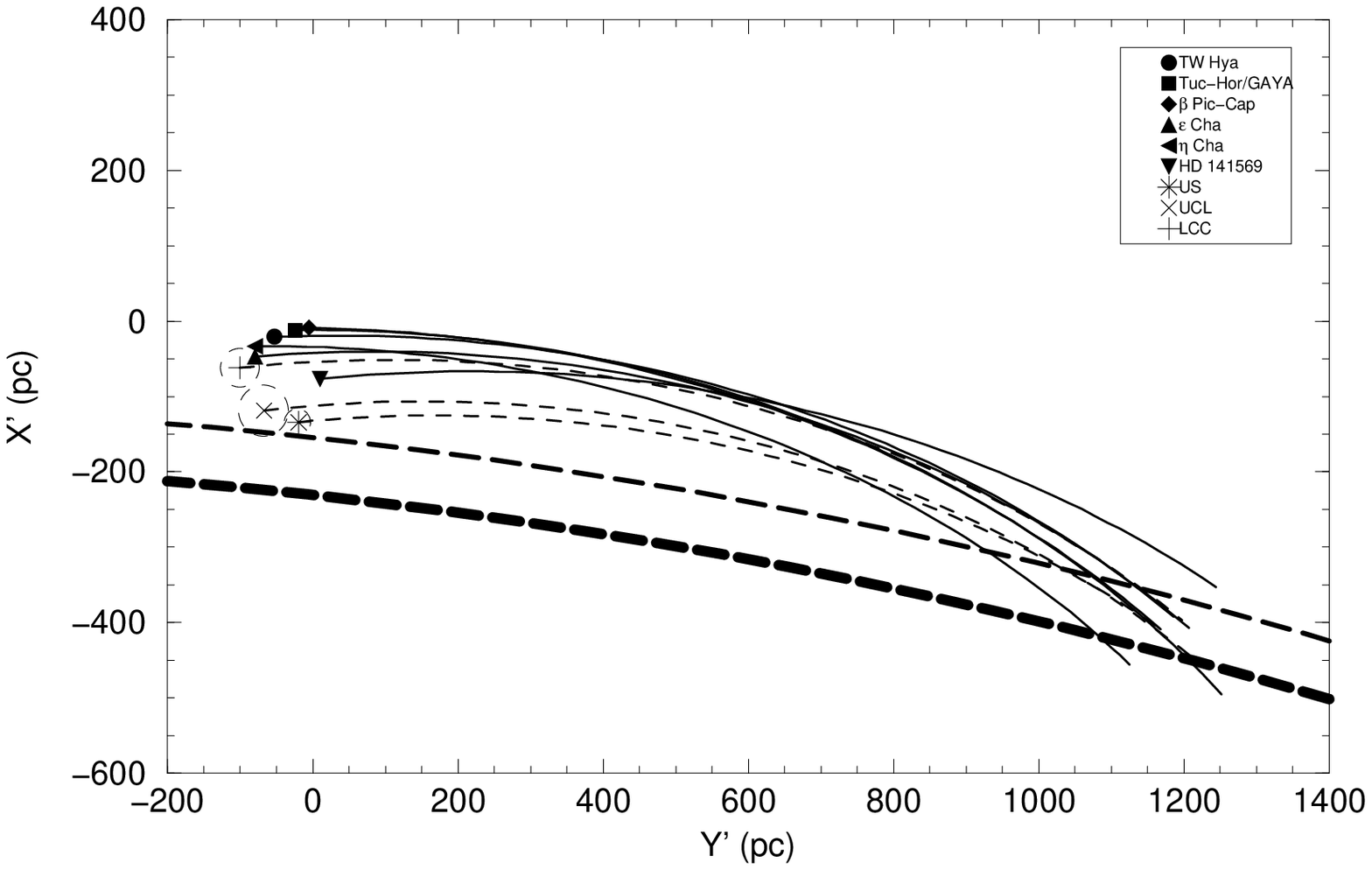}}
\vfill
\vspace{-0.5cm}
\caption{Orbits in the Galactic plane  
         ($X^{\prime}$-$Y^{\prime}$) integrated back in time
         to $t = -$30 Myr for YLA
         and the three associations of the Sco-Cen complex 
         (below, only the orbits for Sco-Cen). The thick-dashed line 
         shows the position of the minimum of the spiral potential 
         (Fern\'andez et al. \cite{Fernandez01}). The thin-dashed line is the 
         position of the phase of the spiral structure $\psi = 10$\degr.}
\label{fig.Associacions+ScoCen+Spiral_xi-eta.30Myr}
\end{figure}

Figure \ref{fig.Associacions+ScoCen+Spiral_xi-eta.30Myr} shows the orbits of
YLA and the OB associations in Sco-Cen for times $0 > t > -30$ Myr in a
galactocentric reference frame ($X'$,$Y'$) which is rotating with an
angular velocity $\Omega = \Omega_\mathrm{p}$ (the rotation velocity of
the spiral arms of the Galaxy; so, in this system the spiral arms remain
fixed). We used the value $\Omega_\mathrm{p} = 30$ km s$^{-1}$ kpc$^{-1}$
obtained by Fern\'andez et al. (\cite{Fernandez01}). The figure shows that,
back in time, the orbits tend to concentrate these associations in a region
with a phase of the spiral structure of 0\degr $\la \psi \la 10$\degr,
very near the spiral potential minimum at $\psi = 0$\degr.

Following the non-linear density wave theory, the motion of the spiral
structure in the Galactic plane generates a shock wave just before the arm
potential minimum for regions outside the corotation radius (Roberts
\cite{Roberts69}; Bertin \& Lin \cite{Bertin96}). In the case of our
galaxy, the phase separation between the position of the shock wave and
the potential minimum is not exactly known (see, for example, Gittins \&
Clarke \cite{Gittins04}). To quantify this, it is common to define the {\it
offset function} $\Theta(R) = m \left( \theta_{\mathrm{shock}} -
\theta_{\mathrm{min}} \right)$, where $m$ is the number of spiral arms of
the Galaxy, and $\theta_{\mathrm{shock}}$ and $\theta_{\mathrm{min}}$ are
the Galactic longitude of the shock wave and the spiral potential minimum
respectively, at galactocentric distance $R$. Roberts (\cite{Roberts69})
derives a small value for this function, close to 0. However, Shu et al.
(\cite{Shu72}) obtain $\Theta(R) = -72$\degr, whereas Yuan \& Grosb\o l
(\cite{Yuan81}) adopt $\Theta(R) = -30$\degr. Despite there being no good
agreement on the value of this parameter, it should have an absolute value
of no more than a few tens of degrees. 

Recent works have studied the formation of a GMC due to the arrival of the 
spiral shock wave, either gathering together pre-existing molecular gas or 
inducing high densities in the shock, converting HI into HII (see Dobbs \& 
Bonnell \cite{Dobbs07} and references herein). According to these authors, GMC 
formation occurs within a very few Myr ($\la5$ Myr) and star formation begins 
very quickly, $\sim$5 Myr after the cloud formation (Clark et al. 
\cite{Clark05}). The first supernovae event occurs about $\sim$4 Myr after 
the formation of the very high mass stars, halting the star formation process.
The stars of the OB association complex are then born, about $\sim$10-15 Myr 
after the arrival of the spiral shock wave.

\begin{figure}
\centering
\resizebox{5cm}{!}{\includegraphics{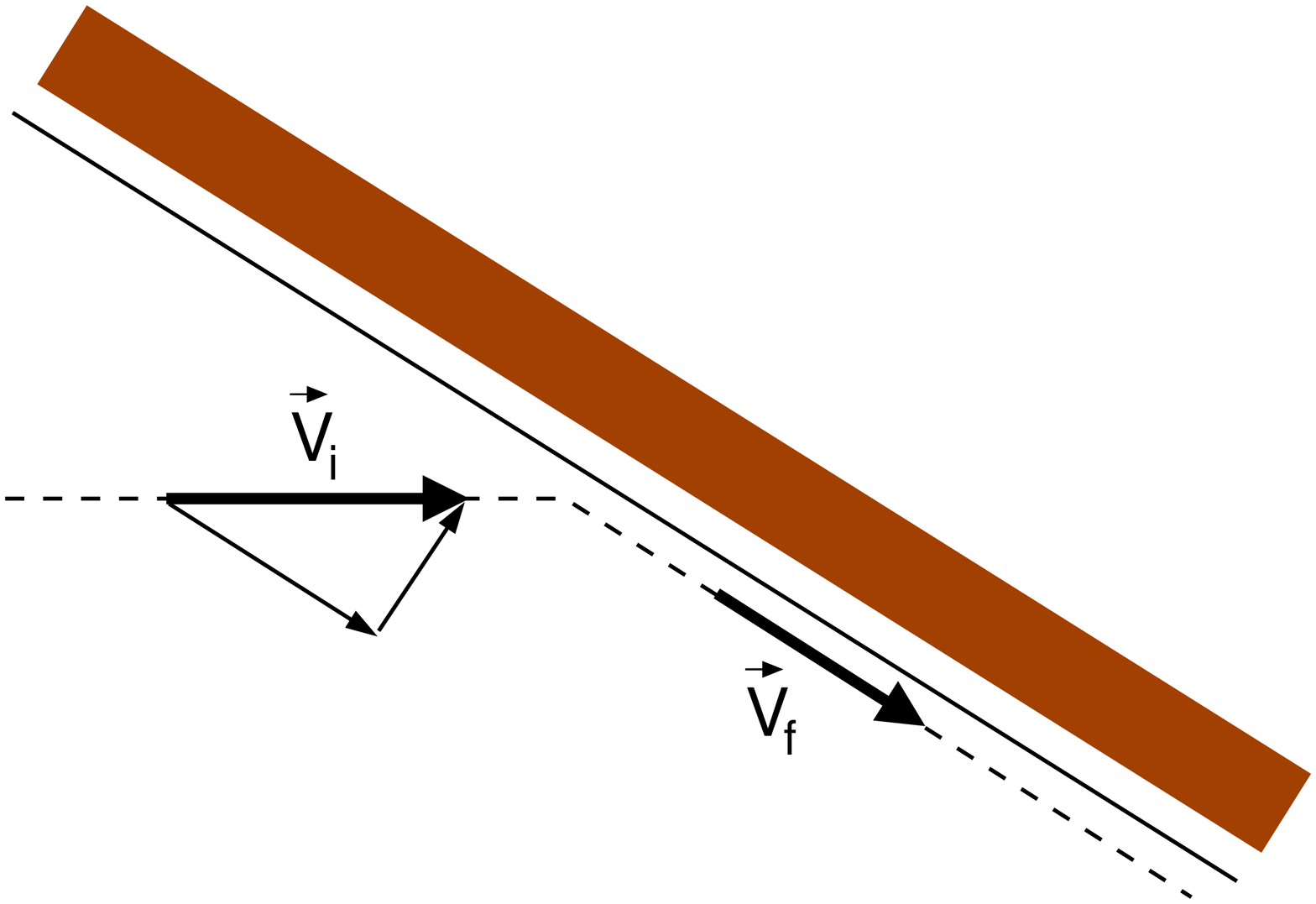}}
\hfill
\leavevmode
\resizebox{5cm}{!}{\includegraphics{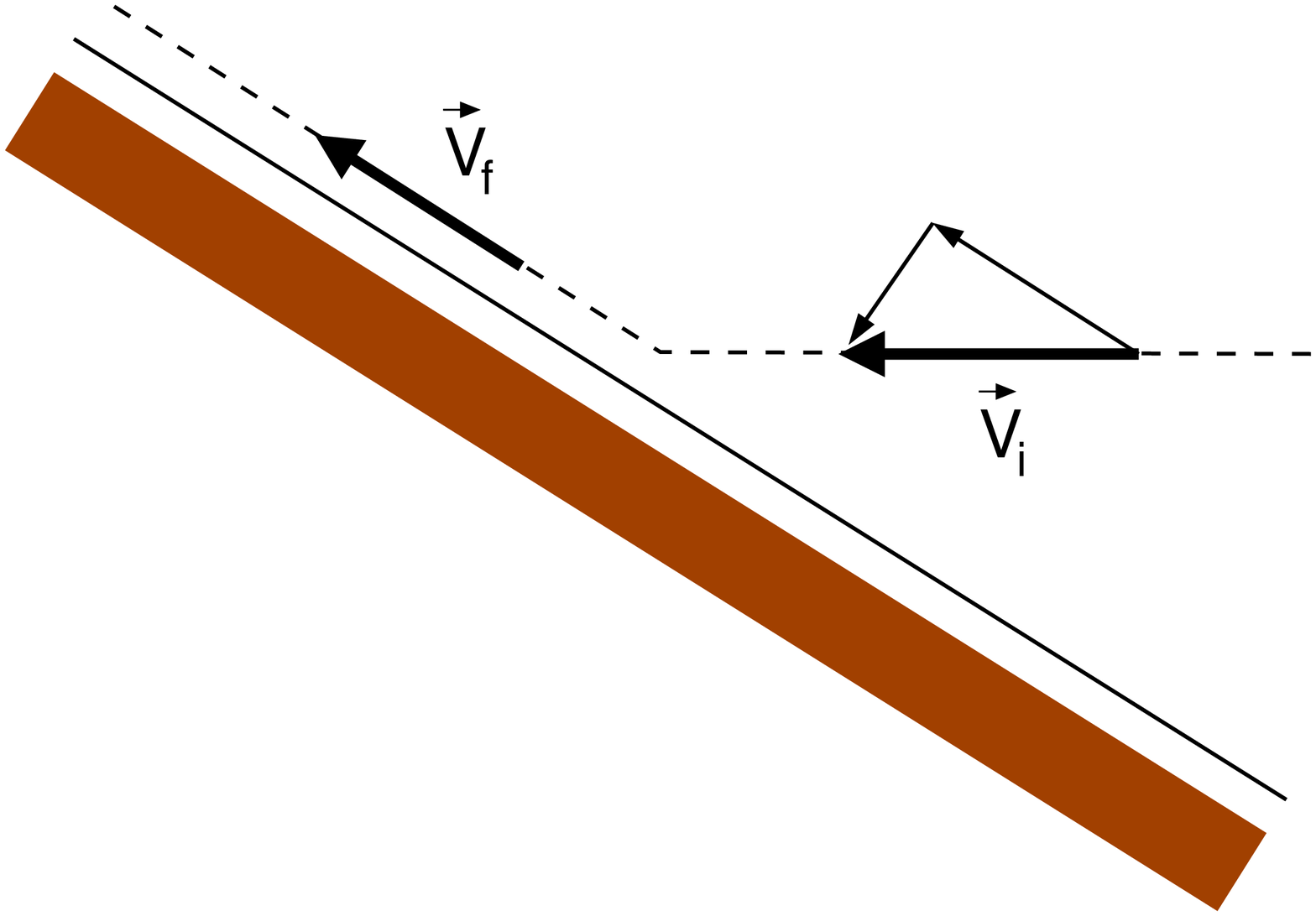}}
\vfill
\caption{Variation of the velocity vector of a test particle when it hits 
         the shock wave (solid line) in the vicinity of the centre of a 
         spiral arm (shadow) and in a region inside (top) and outside (bottom) 
	 the corotation radius (in a reference frame in which the arms are 
	 fixed.)}
\label{fig.OutInCorotation}
\end{figure}

When the shock wave hits the gas, its direction of motion changes.  
Assuming that the shock is strong and sufficiently dissipative, the
velocity component perpendicular to the spiral arm is strongly reduced
(see for example Landau \& Lifshitz \cite{Landau82}).  After interaction
with the shock wave, the gas moves in a direction practically tangential to
the spiral arm (see Fig. \ref{fig.OutInCorotation}). Considering
$\Omega_\mathrm{p} = 30$ km s$^{-1}$ kpc$^{-1}$, in the vicinity of the Sun the
shock wave is moving around 4 km s$^{-1}$ kpc$^{-1}$ faster than the local
standard of rest (LSR; equivalent to 30-35 km s$^{-1}$, depending on the value 
adopted for the distance to Galactic centre) in the Galactic rotation direction. 
In this framework, if the shock wave of a spiral arm hits a molecular cloud, it 
is compressed, and star formation can be triggered.

This is the process proposed by S03 for the formation of the Sco-Cen
complex. They assume that the Sun is placed in a region slightly {\it
inside} the corotation radius. The gas and stars that follow the Galactic
rotational motion are then moving faster than the shock wave. In this
framework, with respect to a reference frame that is rotating at the same
velocity as the spiral arms, the gas is slowed when it hits the shock
wave, and afterwards it moves practically parallel to the spiral arm,
approaching the Galactic centre (see Fig. \ref{fig.OutInCorotation}, top).
However, with respect to LSR, the compressed gas and the newly born stars
have velocities in the direction opposite to the Galactic rotation and
away from Galactic centre. S03 argues that, as this is the {\it present}
motion of the Sco-Cen stars, it favours the following star formation scenario 
for Sco-Cen: a shock with a spiral arm positioned slightly inside the
corotation radius.  However, S03 does not consider the variation in the
orientation of the velocity vector of these stars from their birth to now.

From Fig. \ref{fig.Associacions+ScoCen3_xi-eta} we can see that, in fact,
at their birth the Sco-Cen associations were moving in Galactic
antirotation and {\it away from} Galactic centre. This motion is not
compatible with an interaction with a spiral arm placed inside the
corotation radius, and therefore contradicts the scenario proposed by
S03. However, it is perfectly compatible with the expected velocity if the
arm is outside the corotation radius (see Fig. \ref{fig.OutInCorotation},
bottom), as recently derived in several works (see, for example, Mishurov \&
Zenina \cite{Mishurov99} and Fern\'andez et al. \cite{Fernandez01}). Here,
the difference in velocity between the shock wave and the Galactic
rotation is about 2 km s$^{-1}$ kpc$^{-1}$, so the difference in the
velocity in the direction of Galactic rotation is about 12-13 km s$^{-1}$.
Following numerical simulations of the collapse of nuclei in molecular
clouds (Vanhala \& Cameron \cite{Vanhala98}), one could expect a spiral arm
shock wave to trigger star formation with relative velocities with respect
to the gas of as low as 10 km s$^{-1}$ (though the mechanism is more
efficient when the velocities are between 20 and 45 km s$^{-1}$). This
scenario would then be possible in our case. Moreover, the relative
velocity between the parent molecular cloud and the spiral shock wave will
obviously depend on the cloud's initial velocity with respect to its
regional standard of rest (RSR).

From our orbit calculations, at the moment when the molecular cloud was
generating the first protostars in LCC and UCL (16-20 Myr ago, following
the last estimations published in the literature), the velocity components
of the cloud with respect to its RSR in the Galactic plane were
$(U,V)_{\mathrm{RSR}} \sim (-20,-10)$ km s$^{-1}$. This motion is the
result of the initial proper motion of the cloud and its interaction with
a spiral shock wave ocurring about 30 Myr ago (see Fig. 
\ref{fig.Associacions+ScoCen+Spiral_xi-eta.30Myr}). It is not possible to know 
exactly when this interaction occurred, since we do not know the value of the 
offset function $\Theta(R)$ for the inner arm. However, as we should expect 
small values for this function, we can assume that this interaction took place
approximately at that epoch. This scenario would be in good agreement with the 
present ideas of rapid star formation in molecular clouds that we have 
presented above. The OB associations would have formed about 10-15 Myr after 
the arrival of the spiral shock wave, so about 15-20 Myr ago, in agreement with 
the estimated age for the oldest associations studied here: UCL, LCC and 
Tuc-Hor/GAYA (see Table \ref{tab.Assoc}). In this way, it is possible to 
explain the present observed motion of the Sco-Cen complex, and it is plausible 
that the origin of the star formation in the whole region was an impact with 
the inner spiral arm shock wave.

\subsection{The origin of YLA}\label{sect.originYLA}

Mamajek et al. (\cite{Mamajek00}) found that extrapolating past motions
(assuming linear ballistic trajectories) shows that TW Hya, $\epsilon$
Cha, $\eta$ Cha and the three subgroups of the Sco-Cen complex were
closest together about 10-15 Myr ago. They suggest that these three
associations were formed in the progenitor Sco-Cen GMC or in short-lived
molecular clouds formed by Sco-Cen superbubbles (see also Mamajek \&
Feigelson \cite{Mamajek01}). Jilinski et al. (\cite{Jilinski05}) locate
the birthplaces of $\epsilon$ Cha and $\eta$ Cha at the edge of LCC.  
Ortega et al. (\cite{Ortega02}, \cite{Ortega04}) also suggest that the
$\beta$ Pic moving group was formed near Sco-Cen, probably due to a
supernova in this complex. Our compendium of all the known members of the
whole set of YLA and the integration of their orbits back in time allow us
to present here a more detailed analysis of the origin of YLA.

\begin{figure}
\centering
\resizebox{\hsize}{!}{\includegraphics{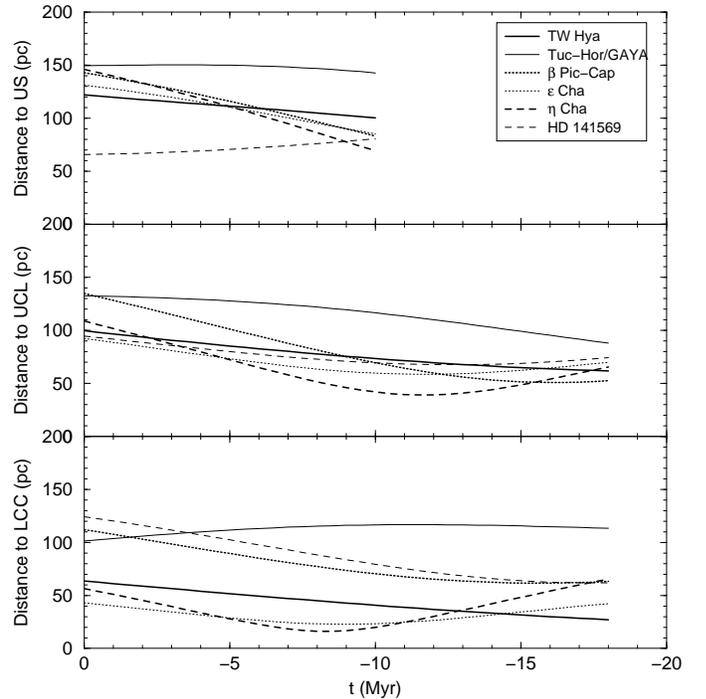}}
\caption{Distances between the centres of YLA
         and the centres of the Sco-Cen associations as a function of 
         time, using the Sco-Cen kinematics in S03.}
\label{fig.DistAss}
\end{figure}

As we show in Sect. \ref{sect.orbits}, the past trajectories of the three
Sco-Cen associations meet in the same region of the first Galactic
quadrant as YLA trajectories do. In Fig. \ref{fig.DistAss} the temporal
evolution of the (three-dimensional) distances between each one of the
local associations and the centres of the three Sco-Cen associations is
shown (with typical estimated errors of a very few tens of pc; see Sect. 
\ref{sect.orbits}). Following these results, LCC is seen to be the association 
closest to YLA in the past.

\begin{table}
   \caption{Distances (in pc) between the centres of YLA and the centres of the 
            Sco-Cen associations at YLA birth. The distance at birth of 
            $\beta$ Pic-Cap to US is not shown because the latter is 
            younger than the former.}
   \label{tab.DistAss}
\centering
\begin{tabular}{lcccccc}
\hline
\hline
Association & \multicolumn{2}{c}{US} & \multicolumn{2}{c}{UCL} & \multicolumn{2}{c}{LCC} \\
\hline
                & M02 & S03 & M02 & S03 & M02 & S03 \\
\hline
TW Hya          &  62 & 104 &  90 &  78 &  90 &  45 \\
$\beta$ Pic-Cap & --- & --- &  80 &  58 & 122 &  65 \\
$\epsilon$ Cha  &  47 &  85 &  75 &  59 &  78 &  23 \\
$\eta$ Cha      &  48 &  69 &  56 &  42 &  60 &  20 \\
HD 141569       &  43 &  70 &  85 &  80 & 120 & 102 \\
\hline
\end{tabular}
\end{table}

The instant when the distance minima between LCC and YLA occurred is of
great interest. In the cases of the $\eta$ Cha cluster and the $\epsilon$
Cha association, minima with distances of 16 and 23 pc, respectively, are
obtained for $t \sim -$(8.5-9) Myr. In the case of the $\beta$ Pic-Cap
moving group, the minimum distance to LCC was 62 pc 15.5 Myr ago. No clear
distance minima to LCC are found in the recent past ($t > -20$ Myr) for
the other YLA. TW Hya was 45 pc from LCC 8 Myr ago (the estimated age for
this association), whereas the HD 141569 system was about 102 pc from LCC 5
Myr ago, continuously decreasing to 62 pc 18 Myr ago. The Tuc-Hor/GAYA
association has maintained a distance to LCC of more than 100 pc over the
last 20 Myr, but this could be the only YLA considered here not to
originate from a supernova in LCC, since its estimated age is equal to
(or even larger than) that derived for LCC.

\begin{figure}
\centering
\resizebox{\hsize}{!}{\includegraphics{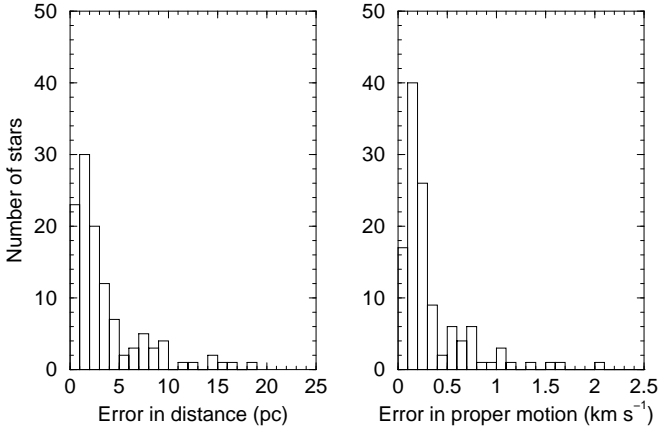}}
\caption{Error in distance (left) and proper motion (right)
         for stars with Hipparcos parallaxes belonging to YLA.}
\label{fig.ErrorR}
\end{figure}

We conclude that, at the moment of their birth, YLA were
at distances of between 20 and 100 pc from the {\it centre} of LCC, and
even further from the other two Sco-Cen associations (see Table
\ref{tab.DistAss}). Although observational errors in parallax and velocity
components, as well as errors in age estimations, could affect these
results, the fact that we work with mean values for distances, velocity
components and ages minimises this possibility. Moreover, even considering
the individual stars, as they are very close, the error in the
trigonometric distances derived from Hipparcos parallaxes is low for most
of them (smaller than 10 pc for 89\% of the stars; see Fig.
\ref{fig.ErrorR}, left). Similarly, the error in proper motions is smaller
than 1.0 km s$^{-1}$ for 92\% of the stars (see Fig. \ref{fig.ErrorR},
right) and errors in radial velocity of less than 2.0 km s$^{-1}$ are
estimated for most of the stars considered here (see Appendix
\ref{sec.app}).

Although the present radius of LCC is about 25 pc, it has been
continuously expanding since birth, 16-20 Myr ago. Even considering a
very moderate expansion rate, one should expect an initial radius $\la$20
pc. The distances obtained in Fig. \ref{fig.DistAss} and presented in
Table \ref{tab.DistAss}, together with the expected reliability of the
orbits back in time (taking into account the low associated errors and the
short integration time interval), lead us to believe that the local 
associations were not born inside the cloud that formed the Sco-Cen complex, 
but in small molecular clouds outside it.

One possible scenario for the formation of YLA in these small molecular clouds 
is the explosion of one or several close supernovae, which could have produced 
compression that triggered star formation. (Lyman continuous photons from a 
luminous O star could also produce triggered star formation: see Lee 
\cite{Lee07}.) These hypothetical supernovae should belong to the Sco-Cen 
complex. As we have seen in Sect. \ref{subsect.ScoCen}, this complex is made up 
of several thousand stars, more than 300 of which are early-type stars, and 
around 35 are candidates for Type II supernovae. (They have spectral types 
between O and B2.5, and masses greater than 8 M$_{\sun}$; see Weiler \& Sramek 
\cite{Weiler98} and Z99.) Ma\'{\i}z-Apell\'aniz (\cite{MaizApellaniz01}) 
estimated the number of past supernovae inside the three associations from the 
Starbust99 models (Leitherer et al. \cite{Leitherer99}), obtaining 1 supernova 
for US, 13 for UCL and 6 for LCC. The first supernova that exploded in each 
association took place when it was 3-5 Myr old, and the others have been 
exploding and will continue to explode at a nearly constant rate, for the first 
$\sim$30 Myr of the complex's life. Even a conservative estimate (considering 
the age for the complex proposed by de Geus et al. \cite{Geus92}) gives at 
least 6 supernovae in UCL during the last 10-12 Myr, another 6 in LCC during 
the last 7-9 Myr and at least 1 in US (see comments in Ma\'{\i}z-Apell\'aniz
\cite{MaizApellaniz01}). There is direct observational evidence for this
(Hoogerwerf et al. \cite{Hoogerwerf01}). If the complex is slightly older
(especially UCL and LCC) as stated by S03 among others, the number of
supernovae in UCL and LCC in the past would increase by 5-10. It is clear that 
there have been several recent supernovae in the solar neighbourhood, although 
it is necessary to know where the Sco-Cen associations were in the past to see 
if they can be linked to the origin of YLA presented in Sect. \ref{subsec.YLA}. 
Figures \ref{fig.Associacions+ScoCen3_xi-eta} and
\ref{fig.Associacions+ScoCen3_xi-eta-zeta} also show the orbits of US, UCL
and LCC back to their birth epochs.

The wave front of a supernova typically moves at a velocity of a few tens
of pc per million years\footnote{After considerable slowing during the
first 3-5 pc of distance covered, the wave front of a typical supernova
moves at $\sim$15-45 km s$^{-1}$ when it is at a distance of $\sim$10-100
pc from the explosion site (see Vanhala \& Cameron \cite{Vanhala98} and
Preibisch et al. \cite{Preibisch02}). A velocity of 30 km s$^{-1}$ is
equal to 30.7 pc Myr$^{-1}$.}; therefore, a supernova explosion in LCC or UCL 
9-11 Myr ago may have triggered star formation between 1 and 3 Myr later in
small molecular clouds at distances of 15 to 75 pc. These would be the
parent clouds of the $\eta$ Cha cluster and the $\epsilon$ Cha
association, located at $\sim$20 pc from LCC at their birth. Taking into
account that the first supernovae in LCC and UCL exploded when the
associations were 3-5 Myr old (Ma\'{\i}z-Apell\'aniz
\cite{MaizApellaniz01}), this scenario is only possible for a present age
of LCC and UCL of at least 12 Myr. This is not a problem, since the
estimated ages for LCC published in the literature range from 11-12 Myr to
16-20 Myr, and for UCL, from 14-15 to 16-20 Myr, as we have seen. In this
way, if only one supernova could explain this star formation outbreak
$\sim$8.5-9 Myr ago, this would be the age of $\eta$ Cha and $\epsilon$
Cha, which would have been formed simultaneously. It should be remembered
that the estimated ages for $\eta$ Cha and $\epsilon$ Cha are 5-15 and $\la$10 
Myr, respectively (see Table \ref{tab.Assoc}).

Such a supernova could also have triggered the formation of TW Hya, whose
estimated age is $\sim$8 Myr. As mentioned above, at that time TW Hya was
about 45 pc from the centre of LCC, in perfect agreement with the typical
distance at which a supernova wave front can trigger star formation in a
small molecular cloud. However, it was not necessarily a single supernova
in LCC or UCL that was the origin of these four YLA. The supernova rate in
these two associations is $\sim$0.5 Myr$^{-1}$ and, therefore, it is
possible that a few supernovae in the period $-8 \la t \la -10$ Myr
triggered the star formation that resulted in YLA. In any case, from our
results we can conclude that these associations definitely did not form
inside LCC or UCL, to be later ejected. They were formed in regions of
space far from LCC and UCL, probably in small molecular clouds that
were later totally dispersed by the newly born stars and/or by the shock
fronts of later supernovae in LCC or UCL.

Our results support a star formation scenario for very young stars far
away from SFR or molecular clouds, such as that proposed by Feigelson
(\cite{Feigelson96}) and not that of Sterzik \& Durisen
(\cite{Sterzik95}). The latter authors perform numerical simulations to
explain the existence of haloes of isolated T Tau stars around SFR. In
their simulations a significant number of stars were ejected from these
regions at birth with large velocities, allowing trajectories of some tens
of pc in a few million years. Meanwhile, Feigelson (\cite{Feigelson96})
proposed another scenario for the formation not only of the haloes of T
Tau stars, but also of other completely isolated very young stars that
have been discovered. In this model, the isolated T Tau stars form in
small, fast-moving, short-lived molecular clouds. The gas remaining after
the star formation process is rapidly dispersed by the stellar winds of
the new stars. At present the stars are located in regions of space where
there is no gas and so, apparently, they have formed far away from any
SFR. The case of YLA supports this scenario, since our kinematic study
shows that these associations formed far away from the Sco-Cen complex. For 
the HD 141569 system, a supernova in UCL, as opposed to one in LCC, is a more 
promising candidate to explain its origin. This is because the distance to LCC 
for the range of ages accepted for this group (2-8 Myr) is between 88 and 116 
pc, whereas for UCL it is 73-88 pc.

\subsection{YLA and LB} \label{sect.YLA+LB}

Several models have been presented to explain the origin of the LB. Cox
(\cite{Cox98}) reviews five conceptions of the LB on which models have been
based. Most of them involve one or more supernovae. The consensus
is reached with a scenario in which about 10-20 supernovae formed the
local cavity and, after that, a few supernovae reheated the LB a few Myr ago,
explaining the currently observed temperature of the SXRB (Breitschwerdt
\& Cox \cite{Breitschwerdt04}). Some authors have remarked that there is
independent evidence for the occurrence of a close supernova ($\sim$30 pc)
$\sim$5Myr ago (Knie et al. \cite{Knie99}), which could be the best
candidate for reheating the LB. Smith \& Cox (\cite{Smith01}) find that one of
the problems with this theory is the low probability of having 2 or 3
isolated supernovae exploding in the solar neighbourhood in a period of a
few million years. In an attempt to solve this problem,
Ma\'{\i}z-Apell\'aniz (\cite{MaizApellaniz01}) proposed that the 2 or 3
supernovae that reheated the local cavity (and which are therefore
responsible for the LB) could have exploded in LCC, but his results faced some
geometrical problems due to the peripheral situation of LCC with respect
to the LB. Bergh\"ofer \& Breitschwerdt (\cite{Berghofer02}) suggest that the
supernovae came from the B1 moving group found by Asiain et al.
(\cite{Asiain99a}), comprised of 33 stars with spectral types B, A and F.
However, the mean distance to the Sun (and so to the centre of the LB) of its
members is $\sim$135 pc and the trajectory back in time of the B1 group
passes through very peripheral zones of the LB (in fact, the B1 group is
formed by stars in the Sco-Cen complex). 

\begin{figure}
\centering
\resizebox{9cm}{!}{\includegraphics{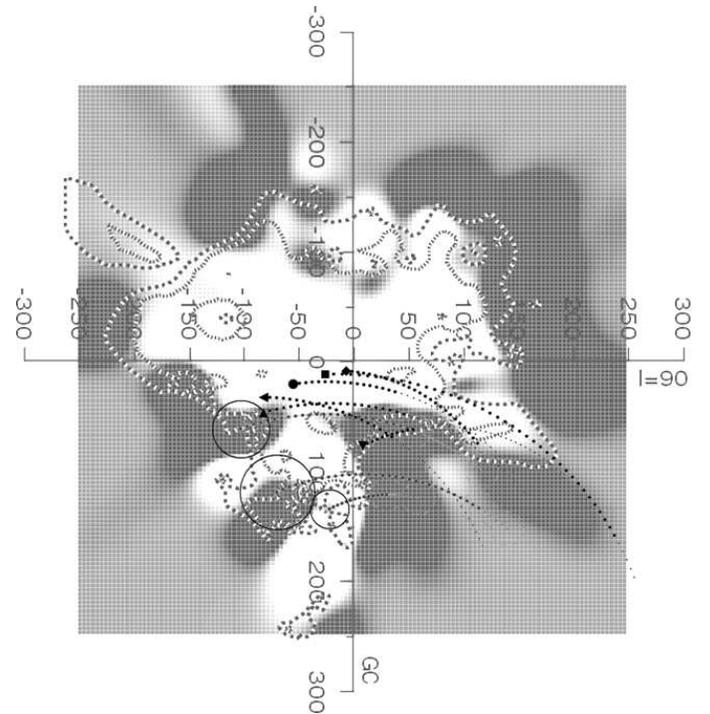}}
\caption{Orbits of YLA and the Sco-Cen associations back in time to their 
         individual ages (in a system comoving with the LSR; see Fig. 
         \ref{fig.Associacions+ScoCen3_xi-eta}) superimposed on the LB 
         structure from Lallement et al. (\cite{Lallement03}) (see Fig. 
         \ref{fig.LB_Lallement}). This structure has been rotated 90\degr 
         with respect to the orientation shown in Fig. 
         \ref{fig.LB_Lallement}.}
\label{fig.LB+Orbits}
\end{figure}

More recently, Fuchs et al. (\cite{Fuchs06}) have proposed that about 14-20 SNe,
originated from the LCC and UCL associations, have been excavating the LB during
the last $\sim$14 Myr. The authors argue that this scenario is realistic even 
though the SNe exploded rather close to the edge of the LB, since the bubble expands
fastest in the direction of the lowest density regions.  In the present case,
the presence of the Loop I in the GC direction would forced the LB to expand towards
the anticentre direction. However, part of the geometric problems that these
works face could be overcome by assuming that one or more recent supernovae
exploded inside YLA. In Fig. \ref{fig.LB+Orbits} we show the trajectories back
in time for YLA, superimposed on the present LB structure derived by 
Lallement et al. (\cite{Lallement03}). Due to small peculiar motions of the
local interstellar gas, the LB is assumed to be nearly at rest in the reference 
frame of the LSR (as done by Fuchs et al. \cite{Fuchs06}). Thus, the position of the
centre of the LB is fixed in time at the system comoving with the LSR (where the 
orbits have been computed; see Fig. \ref{fig.LB+Orbits}). As the LB
has been in expansion during the last Myr, the outer boundaries were placed
nearer to its centre in the past. Due to the uncertainties of the present outer
boundary of the LB and the wealth of data, most models have shortcomings
concerning the dynamical evolution of the structure (Breitschwerdt \& de
Avillez \cite{Breitschwerdt06}). Even when not considering the dynamical
evolution of the LB (i.e., the expansion motion of the boundary), we are able to
obtain important conclusions when comparing the past trajectories of both YLA and
Sco-Cen associations. In Fig. \ref{fig.LB+Orbits} we can see that the past orbits 
of the YLA are closer to the central region of the LB than Sco-Cen 
associations. To 
be exact, the trajectories of the centres of the associations TW Hya, 
Tuc-Hor/GAYA and $\beta$ Pic-Cap have crossed very near to the geometric centre 
of the LB in the last $\sim$5 Myr. As can be seen in the error propagation of 
the kinematic data shown in Fig. 7, the uncertainty boxes in position (due to 
the errors in the mean velocity vector) on the Galactic plane do not exceed a 
few tens of pc. On the other hand, errors in age estimates for the YLA result 
in positional errors along the trajectories back in time. Even considering the 
large uncertainties in age obtained for some of these YLA (see Table 
\ref{tab.AssocAges}), ages are not expected to exceed 20 Myr for any of them 
(except for Tuc-Hor/GAYA). We can therefore conclude that the YLA have been 
moving inside the LB for (at least) most of their lifetime, and can question 
whether the presence of these young stars inside the LB bears any relation to 
its origin and/or evolution.

\begin{table*}
   \caption{Spectral type of known YLA members.}
\label{tab.Asoc.Jov.TS}
\centering
\begin{tabular}{l@{~~}c@{~~}c@{~~}c@{~~}c@{~~}c@{~~}c@{~~}c@{~~}c@{~~}c@{~~}c@{~~}c@{~~}c@{~~}c@{~~}cc}
\hline
\hline
Association     & B2-B4 & B5-B9 & A0-A4 & A5-A9 & F0-F4 & F5-F9 & G0-G4
                & G5-G9 & K0-K4 & K5-K9 & M0-M2 & M3-M5 & M6-M9 &   ?
                & Total \\
\hline
TW Hya          &       &       &   1   &       &       &       &
                &   1   &   2   &   6   &  17   &   3   &   2   &   7
                &  39   \\
Tuc-Hor/GAYA    &   2   &   3   &   5   &   2   &   4   &   9   &   2
                &   5   &   8   &   4   &   2   &   2   &       &   4
                &  52   \\
$\beta$ Pic-Cap &       &       &   3   &   2   &   2   &   5   &   2
                &       &   3   &   4   &   4   &   6   &   1   &   1
                &  33   \\
$\epsilon$ Cha  &       &   1   &   1   &   1   &       &   1   &
                &   1   &   5   &       &   4   &   1   &       &   1
                &  16   \\
$\eta$ Cha      &       &   1   &       &   3   &       &       &
                &       &   3   &       &   4   &   8   &       &
                &  19   \\
HD 141569       &       &   1   &   2   &       &       &       &
                &       &       &       &   1   &   1   &       &
                &   5   \\
Ext. R CrA      &       &   5   &       &   1   &   1   &   1   &   1
                &   4   &  14   &   7   &   8   &  15   &   1   &   1
                &  59   \\
\hline
Total           &   2   &  11   &  12   &   9   &   7   &  16   &   5
                &  11   &  35   &  21   &  40   &  36   &   4   &  14
                & 223   \\
\hline
\end{tabular}
\end{table*}

Table \ref{tab.Asoc.Jov.TS} gives the spectral types of the known members
of the local associations. All the associations except TW Hya and $\beta$
Pic-Cap contain B-type stars (13 stars from a total of 223 members). At
present it is not possible to derive the number of stars earlier than B2.5
that were born in the local associations, since we do not know their
total mass precisely. However, the fact that at present we observe one
supernova candidate ($\alpha$ Pav, a B2IV star belonging to the
Tuc-Hor/GAYA association), and more than a dozen stars of spectral types
between B5 and B9, allows us to affirm that it is possible that one or
more of these associations has sheltered a supernova in the recent past
(the last 10 Myr). As there is direct evidence for an explosion of a
supernova at a distance of $\sim$30 pc, $\sim$5 Myr ago, several
pieces of the same puzzle seem to support the theory of a recent supernova
in the nearest solar neighbourhood originating from a parent star
belonging to a YLA, probably Tuc-Hor/GAYA or the extended R CrA
association (which currently show the highest content of B-type stars).  
This near and recent supernova would have been responsible for the
reheating of the gas inside the LB needed to achieve the currently observed
temperature of the diffuse soft X-ray background.

As we mention above, there is no agreement in the literature on the number
of supernovae needed to form the local cavity. If only one was enough, the 
supernova we propose would be the most promising candidate, since it
would be placed very near the geometric centre of the LB, explaining in a
natural way its present spatial structure. If more supernovae are needed
(as recent works suggest), we could consider other stars in the vicinity
of LCC and UCL, as proposed by Fuchs et al. (\cite{Fuchs06}).

%

\section{Our scenario} 
\label{sect.sce}

If the impact of the spiral arm shock wave was the initial cause of star
formation in the Sco-Cen region (see Sect. \ref{sect.origin.sco-cen}) then
the history of the nearest solar neighbourhood during the last few tens of
millions of years would have been as follows. 30 Myr ago the GMC that
became the parent of Sco-Cen was in the Galactic plane with coordinates
$(X^{\prime},Y^{\prime}) \sim (-400,1200)$ pc (see Fig.  
\ref{fig.Associacions+ScoCen+Spiral_xi-eta.30Myr}). The arrival of the
potential minimum of the inner spiral arm triggered star formation in the
region. At the same time it disturbed the cloud's motion, whose velocity
vector became directed in the opposite direction to Galactic rotation and
away from Galactic centre (just as expected after an interaction with the
spiral arm for a position outside the corotation radius; see Fig.
\ref{fig.OutInCorotation}, bottom). The compression due to the spiral arm
did not necessarily trigger star formation in the whole cloud, but perhaps
only in the regions with the largest densities. This would be favoured by
the smaller relative velocity between the shock wave and the RSR (12-13 km
s$^{-1}$, as we have seen). The regions where star formation began must be
those which generated UCL, LCC and, probably, the Tuc-Hor/GAYA
association, which were all born at nearly the same time: about 16-20 Myr
ago. The stellar winds from the first massive stars began to compress the
gas of the neighbouring regions, maybe causing them to fragment into small
molecular clouds that moved away from the central region of the parent
cloud. About 9 Myr ago, a supernova in LCC or UCL triggered star formation
in these small molecular clouds, giving birth to the majority of YLA, as
we saw in the previous section. The stellar winds of the newly born stars
rapidly expelled the remaining gas from these small clouds (the {\it
cloudlets} proposed by Feigelson \cite{Feigelson96}), completely erasing
every trace of them and leading to our observation that there is no gas in
these regions at present. YLA may have had a crucial influence on the
history of the LB. We suggest that one or two supernovae in these associations
were responsible for reheating the LB a few million years ago. This hypothesis
seems to be reinforced by the evidence of a very near supernova about 5
Myr ago (Knie et al. \cite{Knie99}). At about the same time, as proposed
by Preibisch \& Zinnecker (\cite{Preibisch99}), the shock front of a
supernova in UCL would have triggered star formation in US about 6 Myr
ago. Only 1.5 Myr ago, the most massive star in US would have gone
supernova and its shock front would now be reaching the molecular cloud of
$\rho$ Oph, triggering the beginning of the star formation process there.

%

\section{Conclusions}
\label{sect.conc}

This paper studies the kinematic evolution of the Sco-Cen complex and the
so-called young local associations (YLA). It makes use of most of the
astrophysical data published in the literature for all the known members
of YLA (more than 200 stellar systems). This information appears in
Appendix \ref{sec.app} and can also be accessed via a
webpage\footnote{http://www.am.ub.es/$\sim$dfernand/YLA/}.

The study of the orbits integrated back in time for all these associations
allows us to propose a scenario for recent star formation in the solar
neighbourhood (Sect. \ref{sect.sce}) and a possible link between these 
associations and the origin and/or evolution of the LB (Sect. 
\ref{sect.YLA+LB}). In our scenario, the oldest Sco-Cen associations were 
brought about by the impact of the inner spiral arm against a giant molecular 
cloud. YLA were born later (and outside Sco-Cen) due to the shock fronts of the 
most massive supernovae belonging to the LCC or UCL associations. Our results 
suggest that a YLA is the most likely place to have harboured the supernova 
that reheated the LB a few million years ago.

As seen in this paper, the recent discovery of a set of YLA, together with the 
use of appropriate tools (mainly, the integration back in time of orbits), 
sheds light on several apparently unrelated topics in astrophysics. These 
topics range from star formation mechanisms for low-mass stars distant from 
star forming regions, to the recent history of the LB, and include the origin 
of the Sco-Cen stellar complex and its possible independence from the Gould 
Belt. The possible discovery in the coming years of new members of these 
associations, or even of new associations, will be very useful in confirming 
the results obtained in this work.

\begin{acknowledgements}
This work was supported by the CICYT under contracts AYA2003-07736
and AYA2006-15623-C02-02.
\end{acknowledgements}

%

\begin{appendix}
\section{Young local associations} \label{sec.app}

This Appendix is a compendium of present knowledge about YLA that have
been discovered over the last decade. The information contained in the
tables can be accessed via a
webpage\footnote{http://www.am.ub.es/$\sim$dfernand/YLA/}.

\subsubsection{The TW Hya association}
\label{sect.twhya}

TW Hya has been the prototype of isolated T Tau stars since its classification
as a classical T Tau star (CTTS) of about 10 Myr by Rucinsky \& Krautter
(\cite{Rucinski83}). It is far away from any dark cloud and even from
other PMS stars. Gregorio-Hetem et al. (\cite{Gregorio-Hetem92})  
identified four candidate T Tau stars in a circle of 10\degr\ around TW
Hya. The existence of a real association (the so-called {\it TW Hya
association} or TWA) was finally established when Webb et al.
(\cite{Webb99}) found seven new members, including a brown dwarf. These
new members showed important X-ray emissions, a large abundance of Li,
intense chromospheric activity and a proper motion similar to that
observed in the previously identified members of TWA. Webb et al.
(\cite{Webb99}) derived an age of 8 Myr for the association from the
colour-magnitude diagram. Barrado y Navascu\'es (\cite{Barrado06}), making 
use of the location in the HR diagram and H$\alpha$ and Li equivalent widths 
for individual stars, estimated an age of 10$^{+10}_{-7}$ Myr.

Sterzik et al. (\cite{Sterzik99}), Zuckerman et al. (\cite{Zuckerman01c}),
Makarov \& Fabricius (\cite{Makarov01}), Gizis (\cite{Gizis02}), Reid
(\cite{Reid03}) and Song et al. (\cite{Song03}) added more stars to the
list, increasing the number of members to 25 stellar systems. Table
\ref{tab.TWA} is adapted from Torres et al. (\cite{Torres03}) and shows
the members of TWA, with the addition of the members identified by Reid
(\cite{Reid03}) and Song et al. (\cite{Song03}).

\subsubsection{The Tuc-Hor/GAYA association}
\label{sect.tuchor}

The Tucana and Horologium associations were discovered by Zuckerman \& Webb 
(\cite{Zuckerman00}) and Torres et al. (\cite{Torres00}), independently. The 
former authors performed a search of comoving stars around IRAS sources, 
whereas Torres et al. searched for ROSAT sources around the active star ER Eri. 
Zuckerman et al. (\cite{Zuckerman01a}) proposed merging the two associations 
into one, the so-called {\it Tuc-Hor association}, due to their proximity in 
the sky and the similarities in their velocity components and age. Later, 
Torres et al. (\cite{Torres01}) proposed the denomination {GAYA association} 
for the stars that belonged to the Tuc-Hor association and some other young 
stars in a region of the sky with an angular extent of about 60\degr. Song et
al. (\cite{Song03}) found new members of the association.

Table \ref{tab.TucHor} shows the members of the Tuc-Hor/GAYA association.
The first two sections of the table show the data for the stars originally
assigned to the Tuc association (nucleus and halo of the group). The stars
originally assigned to the Hor association are shown in sections 3 to 5
(probable and possible members, and early-type stars). The sixth section
includes the stars proposed by Zuckerman et al. (\cite{Zuckerman01a}) and
the seventh section lists the stars identified as members by Song et al.
(\cite{Song03}).

\subsubsection{The $\beta$ Pic-Cap moving group}
\label{sect.piccap}

Since Smith \& Terrile (\cite{Smith84}) identified a dust disk around
$\beta$ Pic, determining the age for this A5V star has been a major goal
in the field of planet formation. Barrado y Navascu\'es et al.  
(\cite{Barrado99}) used the list of possible kinematic companions of
$\beta$ Pic to investigate the existence of a real moving group that
could provide clues to the age of this star. Using accurate radial
velocities, proper motions and parallaxes, Barrado y Navascu\'es et al.
(\cite{Barrado99}) found that 6 stars share the motion of $\beta$ Pic, but
only 2 of them have a common age. Zuckerman et al. (\cite{Zuckerman01a})
searched for members of the $\beta$ Pic moving group in the Hipparcos
catalogue and found 18 candidate stellar systems (including some members
previously assigned to the Tucana association by Zuckerman \& Webb
\cite{Zuckerman00} and to the Capricornus association by van der Ancker et
al. \cite{Ancker00}, \cite{Ancker01}) with a mean heliocentric distance of
35 pc. This mean distance meant that $\beta$ Pic was the nearest moving
group to the Sun (though a few years later this honour passed to the AB Dor 
moving group). The age of the group determined by Zuckerman et al.  
(\cite{Zuckerman01a}) is about 12 Myr. Song et al. (\cite{Song03}) added new 
members to the group, which has a spatial extent of about 100 pc.

Ortega et al. (\cite{Ortega02}) studied the orbits back in time of the
members of the $\beta$ Pic-Cap moving group and found a kinematic age of
11.5 Myr. At that time the group was at a distance of about 45 pc from the
centre of the LCC and UCL complexes in Sco-Cen. The authors believe that
an explosion in the periphery of LCC or UCL 12.5 Myr ago could have
triggered the formation of the $\beta$ Pic group 1 Myr later.

Table \ref{tab.BetaPic} lists the known members of the $\beta$ Pic-Cap moving 
group: in the first section, there are those stars identified by Zuckerman et
al. (\cite{Zuckerman01a}) and in the second section, those selected by
Song et al. (\cite{Song03}).

\subsubsection{The $\epsilon$ Cha association}
\label{sect.epscha}

The Chamaeleon cloud complex was first studied by Hoffmeister
(\cite{Hoffmeister62}), who identified several RW Aur-type variable stars,
some of them showing H$\alpha$ emission. Before the ROSAT mission, it was
suspected that 14 stars were connected to the Chamaeleon region. The
optical counterparts of the ROSAT sources identified several dozen new
candidate PMS stars (Alcal\'a et al. \cite{Alcala95}, \cite{Alcala97}).
Thirty of them were confirmed through Li abundance calculations (Covino et
al. \cite{Covino97}).

The so-called {\it $\epsilon$ Cha association} was discovered by Frink et
al. (\cite{Frink98}; see also Frink \cite{Frink99}), studying the
kinematics of T Tau stars located not only in the cloud nuclei, but also
in regions far away from them. Mamajek et al. (\cite{Mamajek00}) added
some stars detected by Terranegra et al. (\cite{Terranegra99}) and Eggen
(\cite{Eggen98}) to the group.

Table \ref{tab.EpsCha} lists the known members of the $\epsilon$ Cha
association, with data from Frink et al. (\cite{Frink98}) Terranegra et
al. (\cite{Terranegra99}) and Mamajek et al. (\cite{Mamajek00}).

\subsubsection{The $\eta$ Cha cluster}
\label{sect.etacha}

The $\eta$ Cha cluster was discovered by Mamajek et al.  
(\cite{Mamajek99}) from observations made in 1997 using the High Resolution 
Imager (HRI) of the ROSAT satellite. ROSAT had previously detected
four X-ray sources around the star $\eta$ Cha, which were identified as
weak-line T Tau stars (Alcal\'a et al. \cite{Alcala95}; Covino et al. 
\cite{Covino97}). From the HRI observations, 12 X-ray sources were discovered 
in an area of 0.2 sqr deg, each one with a prominent optical counterpart. 
Spectroscopic observations of these stars were made by Mamajek et al. 
(\cite{Mamajek99}) and led to the stars being classified as PMS stars 
belonging to the $\eta$ Cha cluster. The brightest stars in the cluster have 
similar parallaxes and proper motions in the Hipparcos catalogue, confirming 
the cluster's identity. Lawson (\cite{Lawson01}) and Lawson et al. 
(\cite{Lawson02}) discovered two new members of the cluster, while Song et 
al. (\cite{Song04}), Ran Lyo et al. (\cite{RanLyo04}) and Luhman \& Steeghs 
(\cite{Luhman04}) independently added three new members, giving a total of 18 
stellar systems with 9 secondary stars.

Table \ref{tab.EtaCha} lists the known members of the $\eta$ Cha cluster.
It should be complete down to 0.15 M$_{\sun}$ (or spectral type M6; see
Ran Lyo et al. \cite{RanLyo04}) or even 0.015 M$_{\sun}$ (Luhman \&
Steeghs \cite{Luhman04}).

\subsubsection{The HD 141569 system}
\label{sect.hd}

Weinberger et al. (\cite{Weinberger01}) searched around the triple star HD
141569 Hipparcos for candidates sharing distance and motion with this
system. They found that two stars of spectral type A lie below or along
the ZAMS, where the young A-type stars are placed.

Table \ref{tab.HD141569} lists the members of the HD 141569 system; a
total of 3 stellar systems with 5 stars.

\subsubsection{The extended R CrA association}
\label{sect.rcra}

In the solar neighbourhood, the densest nucleus of a molecular cloud is
the dark cloud near the star R CrA (Dame et al. \cite{Dame87}), which has
an extinction of up to 45 magnitudes in the visible range. Several IR
surveys have shown the existence of a large population of IR sources
inside the cloud (see, for example, Taylor \& Storey \cite{Taylor84}, or
Wilking et al. \cite{Wilking97}). The age of the cloud has been estimated
at between 1 Myr (Knacke et al. \cite{Knacke73}) and 6 Myr (Wilking et al.  
\cite{Wilking92}). Marraco \& Rydgren (\cite{Marraco81}) obtained a
distance of about 129 pc for the cloud. The large error in the Hipparcos
parallax for the star R CrA did not shed new light on this issue.

Since the 1970s, a few classical T Tau stars have been associated with the
CrA dark cloud (see for example Knacke et al. \cite{Knacke73}; Wilking et
al. \cite{Wilking86}; Wilking et al. \cite{Wilking92}). More recently,
Neuh\"auser et al. (\cite{Neuhauser00}) conducted an optical
identification program between the non-identified RASS (ROSAT All-Sky
Survey) sources to find PMS stars in and around the CrA dark cloud. They
found 19 optical counterparts that fulfil PMS star conditions; 2 of them
are classical T Tau stars, whereas the other stars are weak-line T Tau
stars. The spatial distribution of these stars does not show the
proximity to the cloud observed in the former members.  Another 21
possible members were found by Quast et al. (\cite{Quast01}) in a region
of the sky with a projected diameter of around 35\degr. This fact led the
authors to rename the group {\it extended R CrA association}.

Table \ref{tab.RCrA} shows the known members of the extended R CrA
association. The two first sections include those PMS stars known before
the ROSAT mission and 4 B-type stars that could also be associated with
the CrA cloud. The third section lists the 19 new members, all of them T
Tau stars, discovered by Neuh\"auser et al. (\cite{Neuhauser00}). The last
section shows the two new members explicitly mentioned by Quast et al.
(\cite{Quast01}) among the 21 new members that they found.

\subsubsection{The AB Dor moving group}
\label{sect.abdor}

Zuckerman et al. (\cite{Zuckerman04a}) identified about 40 nearby stars
that are moving through space together with the intensively studied star
AB Dor. The mean distance to these stars is 32 pc, making it the closest
group to Earth. The authors performed spectroscopic observations and
derived radial velocities, rotational velocities and equivalent widths of
the H$\alpha$ and Li $\lambda$6708 lines. They estimated the age of the
moving group as $\sim$50 Myr, comparing the H$\alpha$ emission and
absorption of the late K- and early M-type stars in the AB Dor moving
group with those identified in the Tucana association. Luhman et al.  
(\cite{Luhman05}) estimated an age in the range 75-150 Myr using
colour-magnitude diagrams. This would make the moving group roughly coeval
with the Pleiades ($\tau \sim$ 100-125 Myr; see, for example, Stauffer et
al. \cite{Stauffer98} and Meynet et al. \cite{Meynet93}) and therefore
much older than the other associations studied in this section. However,
L\'opez-Santiago et al. (\cite{Lopez06}) found that the moving group can
be split into two subgroups, with ages of 30-50 Myr and 80-120 Myr. On
the other hand, Makarov (\cite{Makarov07}) proposes that the nucleus of the
AB Dor moving group formed 38 Myr ago during a close passage of the
Cepheus OB6 cloud. In any case, we can see that an age greater than 30 Myr
is proposed in all works.

Table \ref{tab.ABDor} shows the members of the AB Dor moving group found
by Zuckerman et al. (\cite{Zuckerman04a}). The first section of the table
includes the more probable members, and the second, shows stars whose 
membership of the moving group is questionable.

\end{appendix}

%
%

%
%

\longtabL{7}{
\begin{landscape}
\begin{longtable}{@{}l@{~~}l@{~~}cc@{~}c@{~}cc@{}c@{}c@{}c@{}c@{}c@{}}
   \caption{Known members of the TW Hya association. See details in Sect. 
            \ref{sect.twhya}.}\\
\hline
\hline
 & & & & & & & $\mu_\alpha \cos \delta$ & $\mu_\delta$ & 
 $\pi_{\mathrm {Hip}}$ & Eq. width & Mean $V_R$ \\
 TWA & Other denominations & R.A. (J2000) & Dec. (J2000) &
 Sp.T. & V & B-V & (mas yr$^{-1}$) & (mas yr$^{-1}$) & 
 (mas) & Li $\lambda$6708 & (km s$^{-1}$) \\
 & & & & & & & & & & (\AA) & \\
\hline
TWA-1   &  TW Hya                   &  11h 01m 51.9s  &  $-34\degr 42\arcmin 17\arcsec$  &  K7e     &     10.92               &  0.97 &  ~$-$73.4~$\pm$~~2.3 &  $-$17.5~$\pm$~~2.3 &  17.72~$\pm$~2.21 &  0.426         &  $+$12.66~$\pm$~0.22 \\
TWA-2A  &  CD$-29\degr$8887A        &  11h 09m 13.8s  &  $-30\degr 01\arcmin 40\arcsec$  &  M2e     &     11.07               &  1.42 &  ~$-$90.1~$\pm$~~4.3 &  $-$21.1~$\pm$~~4.0 &  19.23~~~~~~~~~~  &  0.494         &  $+$10.97~$\pm$~0.12 \\
TWA-2B  &  CD$-29\degr$8887B        &    $+0\farcs3$  &                     $+0\farcs5$  &  M2      & ~\llap{$+$}0.81\rlap{K} &       &                      &                   \ &                 \ &  \             &  \                   \\
TWA-3A  &  Hen~3-600A               &  11h 10m 28.0s  &  $-37\degr 31\arcmin 52\arcsec$  &  M3e     &     12.04               &  1.52 &  ~$-$96.5~$\pm$~11.8 &  $+$16.3~$\pm$~11.9 &  23.81~~~~~~~~~~  &  0.563         &  $+$11.2~~~~~~~~~~~~ \\
\end{longtable}
\end{landscape}
}

%
%

\longtabL{8}{
\begin{landscape}
\begin{longtable}{l@{~~}l@{~~}c@{~~}c@{~~}c@{~~}c@{~~}c@{~~}c@{~~}c@{~~}c@{~~}c@{~~}c@{~~}c}
   \caption{Probable and possible members of the Tuc-Hor/GAYA association. 
            See details in Sect. \ref{sect.tuchor}.} \\
\hline
\hline
 & & & & & & & $\mu_\alpha \cos \delta$ & $\mu_\delta$ & 
 $\pi_{\mathrm {Hip}}$ & Eq. width & Mean $V_R$ \\
 Star & Other den. & 
 R.A. (J2000) & Dec. (J2000) & Sp.T. & V & 
 B-V & (mas yr$^{-1}$) & (mas yr$^{-1}$) & (mas) & 
 Li $\lambda$6708 & (km s$^{-1}$) \\
 & & & & & & & & & & (\AA) & \\
\hline
HIP 1481      &                   & 00h 18m 26.0s & $-63\degr 28\arcmin 38\arcsec$ & F8        & ~7.45 &   ~0.54         & ~$+$90.37~$\pm$~~0.67 & ~$-$58.98~$\pm$~~0.60 & 24.42~$\pm$~~0.68 &           0.125 & ~$+$7.0~$\pm$~~2.0 \\
HIP 1910      &                   & 00h 24m 08.9s & $-62\degr 11\arcmin 04\arcsec$ & M0        & 11.47 &   ~1.39         & ~$+$92.71~$\pm$~~2.49 & ~$-$45.92~$\pm$~~1.73 & 21.59~$\pm$~~2.22 &           0.210 & ~$+$4.0~$\pm$~~2.0 \\
HIP 1993      &                   & 00h 25m 14.6s & $-61\degr 30\arcmin 48\arcsec$ & K7        & 11.47 &   ~1.35         & ~$+$84.59~$\pm$~24.19 & ~$-$63.40~$\pm$~18.94 & 26.69~$\pm$~24.06 & \llap{$<$}0.50~~& ~$+$7.0~$\pm$~~2.0 \\
HIP 2484      &                   & 00h 31m 32.6s & $-62\degr 57\arcmin 29\arcsec$ & B9        & ~4.38 & ~\llap{$-$}0.07 & ~$+$82.48~$\pm$~~0.47 & ~$-$54.37~$\pm$~~0.42 & 23.35~$\pm$~~0.52 & ...             & ~$+$9.0~$\pm$~10.0 \\
\end{longtable}
\end{landscape}
}

%
%

\longtabL{9}{
\begin{landscape}
\begin{longtable}{ll@{~~}c@{~~}c@{~~}cccccc@{~~}c@{~~}c}
   \caption{Known members of the $\beta$ Pic moving group. See details in 
            Sect. \ref{sect.piccap}.}\\
\hline
\hline
 & & & & & & & $\mu_\alpha \cos \delta$ & $\mu_\delta$ &
 $\pi_{\mathrm {Hip}}$ & Eq. width & $V_R$ \\
 Star & Other den. & R.A. (J2000) & Dec. (J2000) & Sp.T. & V & V-I & 
 (mas yr$^{-1}$) & (mas yr$^{-1}$) & (mas) & 
 Li $\lambda$6708 & (km s$^{-1}$) \\
 & & & & & & & & & & (\AA) & \\
\hline
HIP 560     & HR 9                        & 00h 06m 50.0s & $-23\degr 06\arcmin 27\arcsec$ & F2IV     & ~6.19 &           0.46 & ~$+$97.30~$\pm$~0.73 & ~$-$47.09~$\pm$~0.42 & ~25.59~$\pm$~0.75 &                 & $+$15.9~~$\pm$~0.5~ \\
HIP 21547   & 51 Eri                      & 04h 37m 36.1s & $-02\degr 28\arcmin 24\arcsec$ & F0V      & ~5.22 &           0.34 & ~$+$43.32~$\pm$~0.81 & ~$-$64.23~$\pm$~0.61 & ~33.60~$\pm$~0.91 &                 & $+$21.0~~$\pm$~2.~~ \\
            & GJ 3305                     & 04h 37m 37.3s & $-02\degr 29\arcmin 37\arcsec$ & M0.5     & 10.59 &           1.90 &                      &                      &                   &                 &                     \\
HIP 23309   &                             & 05h 00m 47.1s & $-57\degr 15\arcmin 26\arcsec$ & M0/1     & 10.01 &           1.79 & ~$+$35.64~$\pm$~1.10 & ~$+$72.80~$\pm$~0.97 & ~38.08~$\pm$~1.07 &           0.294 & $+$17.8~~$\pm$~0.8~ \\
\end{longtable}
\end{landscape}
}

%
%

\longtabL{10}{
\begin{landscape}
\begin{longtable}{ll@{~~}c@{~~}c@{~~}ccccc@{~~}c@{~~}c}
   \caption{Known members of the $\epsilon$ Cha association. See details 
            in Sect. \ref{sect.epscha}.}\\
\hline
\hline
 & & & & & & & $\mu_\alpha \cos \delta$ & $\mu_\delta$ &
 $\pi_{\mathrm {Hip}}$ & Mean $V_R$ \\
 Star & Other denominations & R.A. (J2000) & Dec. (J2000) & Sp.T. & V & B-V & 
 (mas yr$^{-1}$) & (mas yr$^{-1}$) & (mas) & (km s$^{-1}$) \\
\hline
\hline
EG Cha          & 1RXS J083656.7-785643       & 08h 36m 56.2s & $-78\degr 56\arcmin 46\arcsec$ & K1       & 10.68 &           1.34  &  $-$29.5~~$\pm$~2.1~ & $+$27.6~~$\pm$~1.6~ &                  & $+$18.~~~~~~~~~~~ \\
VW Cha          & RX J1108.1-7742             & 11h 08m 01.8s & $-77\degr 42\arcmin 29\arcsec$ & ?e...    & 13.36 &           1.24  &  $-$46.~~~$\pm$~8.~~ &  $+$8.~~~$\pm$~8.~~ &                  & \\
RX J1150.4-7704 & GSC 09415-01685             & 11h 50m 28.9s & $-77\degr 04\arcmin 38\arcsec$ & K4       & 12.0~~&                 &  $-$40.2~~$\pm$~5.1~ &  $-$5.8~~$\pm$~1.3~ &                  & \\
RX J1150.9-7411 &                             & 11h 50m 54.8s & $-74\degr 11\arcmin 27\arcsec$ & M4       & 14.4~~&                 &  $-$39.0~~$\pm$~4.2~ &  $+$4.4~~$\pm$~2.9~ &                  & \\
T Cha           & HIP 58285, RX J1157.2-7921  & 11h 57m 13.5s & $-79\degr 21\arcmin 32\arcsec$ & F5       & 11.86 &                 &  $-$39.61~$\pm$~5.08 &  $-$9.87~$\pm$~3.79 & 15.06~$\pm$~3.31 & $+$17.4~$\pm$~1.5 \\
\end{longtable}
\end{landscape}
}

%
%

\longtabL{11}{
\begin{landscape}
\begin{longtable}{l@{~~}l@{~~}c@{~~}c@{~~}c@{~~}c@{~~}c@{~~}c@{~~}c@{~~}c@{~~}c@{~~}c}
   \caption{Known members of the cluster $\eta$ Cha. See details in 
            Sect. \ref{sect.etacha}.}\\
\hline
\hline
 & & & & & & & $\mu_\alpha \cos \delta$ & $\mu_\delta$ &
 $\pi_{\mathrm {Hip}}$ & Eq. width & $V_R$ \\
 Star & Other den. & R.A. (J2000) & Dec. (J2000) & Sp.T. & V & V-I & 
 (mas yr$^{-1}$) & (mas yr$^{-1}$) & (mas) & 
 Li $\lambda$6708 & (km s$^{-1}$) \\
 & & & & & & & & & & (\AA) & \\
\hline
ECHA J0836.2-7908 &                               & 08h 36m 10.6s & $-79\degr 08\arcmin 18\arcsec$ & M5.5      & 17.66 & 3.76 &                      &                     &                  & 0.6~~&                   \\
RECX 1            & EG Cha                        & 08h 36m 56.2s & $-78\degr 56\arcmin 46\arcsec$ & K4        & 10.61 & 1.42 & $-$29.5~~~$\pm$~2.1~~&$+$27.6~~~$\pm$~1.6~ &                  & 0.52 & $+$18.~~$\pm$~2.~ \\
ECHA J0838.9-7916 &                               & 08h 38m 51.5s & $-79\degr 16\arcmin 14\arcsec$ & M5.2      & 16.82 & 3.54 & $-$34.~~~~$\pm$~3.~~~&$+$24.~~~~$\pm$12.~~ &                  & 0.3~~&                   \\
RECX 2            & $\eta$ Cha, HIP 42637         & 08h 41m 19.5s & $-78\degr 57\arcmin 48\arcsec$ & B8V       & ~5.46 &      &  $-$29.31~$\pm$~0.44 & $+$27.58~$\pm$~0.39 & 10.32~$\pm$~0.43 & 0.0~~& $+$14.~~~~~~~~~~  \\
\end{longtable}
\end{landscape}
}

%
%

\longtabL{12}{
\begin{landscape}
\begin{longtable}{ll@{~~}c@{~~}c@{~~}cccccc@{~~}c@{~~}c} 
   \caption{Known members of the HD 141569 system. See details in Sect.
            \ref{sect.hd}.}\\
\hline
\hline
 & & & & & & & $\mu_\alpha \cos \delta$ & $\mu_\delta$ &
 $\pi_{\mathrm {Hip}}$ & Eq. width & $V_R$ \\
 Star & Other den. & R.A. (J2000) & Dec. (J2000) & Sp.T. & V & B-V & 
 (mas yr$^{-1}$) & (mas yr$^{-1}$) & (mas) & 
 Li $\lambda$6708 (\AA) & (km s$^{-1}$) \\
\hline
HD 140574  & HIP 77089 & 15h 44m 26.3s & $-3\degr 50\arcmin 18\arcsec$ & A3     & ~8.46 & 0.39 &  $-$27.16~$\pm$~1.33 &  $-$30.22~$\pm$~0.90 & 10.80~$\pm$~1.15 & ...             & $-$9.4~$\pm$~6.5 \\
HD 141569A & HIP 77542 & 15h 49m 57.7s & $-3\degr 55\arcmin 16\arcsec$ & B9.5Ve & ~7.11 & 0.10 &  $-$16.86~$\pm$~0.98 &  $-$21.11~$\pm$~0.71 & 10.10~$\pm$~0.83 & ...             &$-$6.~~~$\pm$~5.~~\\
HD 141569B &           &       $-$0.5s &                   $-7\arcsec$ & M2V    & 12.0~~& ...  &  ...                 &  ...                 & ...              & 0.50~$\pm$~0.04 & $-$1.5~$\pm$~0.6 \\
HD 141569C &           &       $-$0.5s &                   $-7\arcsec$ & M4V    & 12.0~~& ...  &  ...                 &  ...                 & ...              & 0.50~$\pm$~0.04 & $-$2.4~$\pm$~1.1 \\
HD 141693  & HIP 77570 & 15h 50m 11.9s & $+5\degr 57\arcmin 17\arcsec$ & A0     & ~7.11 & 0.15 &  $-$29.74~$\pm$~1.23 &  $-$26.54~$\pm$~0.70 & ~8.96~$\pm$~0.96 & ...             & ...             \\
\hline
\label{tab.HD141569}
\end{longtable}
\end{landscape}
}

%
%

\longtabL{13}{
\begin{landscape}
\begin{longtable}{l@{~}l@{~}c@{~~}c@{~~}c@{~~}c@{~}c@{~}c@{~}c@{~}c@{~}c@{~}c}
   \caption{Known members of the Extended R CrA association. See details 
            in Sect. \ref{sect.rcra}.}\\
\hline
\hline
 & & & & & & & $\mu_\alpha \cos \delta$ & $\mu_\delta$ &
 $\pi_{\mathrm {Hip}}$ & Eq. width & $V_R$ \\
 Star & Other den. & R.A. (J2000) & Dec. (J2000) & Sp.T. & V & B-V & 
 (mas yr$^{-1}$) & (mas yr$^{-1}$) & (mas) & 
 Li $\lambda$6708 & (km s$^{-1}$) \\
 & & & & & & & & & & (\AA) &  \\
\hline
\hline
S CrA                 & Hen 3-1731       & 19h 01m 08.7s & $-36\degr 57\arcmin 20\arcsec$ & K6     & 10.70 & ...  & $+$10.~~~$\pm$~~~4.6~ & $-$26.~~~$\pm$~~4.6~ & ...                & 0.39 & ~$-$0.~~~~~~~~~~~~ \\
TY CrA                & TYC 7421-1126-1  & 19h 01m 40.8s & $-36\degr 52\arcmin 34\arcsec$ & B9e    & ~9.42 & 0.46 & ~$-$5.30~$\pm$~~~3.50 & $-$30.80~$\pm$~~1.90 & ...                & sí   & ...               \\
R CrA                 & HIP 93449        & 19h 01m 53.7s & $-36\degr 57\arcmin 08\arcsec$ & A5eII  & 11.50 & 0.88 & $-$34.36~$\pm$~102.69 & $+$50.57~$\pm$~40.95 & 121.75~$\pm$~68.24 & ...  & $-$36.0~~~~~~~~~~ \\
DG CrA                & Hen 3-1734       & 19h 01m 55.3s & $-37\degr 23\arcmin 40\arcsec$ & K0     & ...   & ...  & ...                   & ...                  & ...                & 0.57 &                   \\
\end{longtable}
\end{landscape}
}

%
%

\longtabL{14}{
\begin{landscape}
\begin{longtable}{l@{~~~}l@{~~~}c@{~~~}c@{~~~}c@{~~~}c@{~~~}c@{~~~}c@{~~~}c@{~~~}c@{~~~}c@{~~~}c}
   \caption{Known members of the AB Dor moving group. See details in Sect.
            \ref{sect.abdor}.}\\
\hline
\hline
 & & & & & & & $\mu_\alpha \cos \delta$ & $\mu_\delta$ &
 $\pi_{\mathrm {Hip}}$ & Eq. width & $V_R$ \\
 Star & Other den. & R.A. (J2000) & Dec. (J2000) & Sp.T. & V & B-V & 
 (mas yr$^{-1}$) & (mas yr$^{-1}$) & (mas) &
 Li $\lambda$6708 & (km s$^{-1}$) \\
 & & & & & & & & & & (\AA) &  \\
\hline
\hline
           & PW And           & 00h 18m 20.9s & $+30\degr 57\arcmin 22\arcsec$ & K2V    & ~8.6~ & 1.0~ & $+$143.7~~$\pm$~~1.2~ & $-$171.5~~$\pm$~~1.1~ & 45.7~~$\pm$~10.2~ &        & $-$11.0~$\pm$~0.1 \\
HIP 3589   & HD 4277          & 00h 45m 50.9s & $+54\degr 58\arcmin 40\arcsec$ & F8V+K3 & ~7.78 & 0.50 & ~$+$95.62~$\pm$~~1.42 & ~$-$76.91~$\pm$~~0.95 & 20.56~$\pm$~~1.69 &  0.119 & $-$15.4~$\pm$~0.5 \\
HIP 6276   & SAO 147747       & 01h 20m 32.3s & $-11\degr 28\arcmin 04\arcsec$ & (G8)   & ~8.42 & 0.75 & $+$110.69~$\pm$~~1.14 & $-$138.85~$\pm$~~0.56 & 28.52~$\pm$~~1.17 &  0.145 & ~$+$9.9~$\pm$~1.0 \\
HIP 12635  & BD+37 604B       & 02h 42m 20.9s & $+38\degr 37\arcmin 22\arcsec$ & (K3.5) & 10.10 & 0.86 & $+$106.95~$\pm$~16.60 & $-$102.80~$\pm$~12.38 & 22.27~$\pm$~~8.29 &  0.146 & ~$-$4.1~$\pm$~0.3 \\
\end{longtable}
\end{landscape}
}
\typeout{get arXiv to do 4 passes: Label(s) may have changed. Rerun}

\begin{thebibliography}{}

\bibitem[1995]{Alcala95} 
Alcal\'a, J.M. Krautter, J., Schmitt, J.H.M.M., Covino, E., Wichmann, R., 
Mundt, R. 1995, \aaps, 114, 109

\bibitem[1997]{Alcala97} 
Alcal\'a, J.M. Krautter, J., Covino, E., Neuh\"auser, R., Schmitt, 
J.H.M.M., Wichmann, R. 1997, \aap, 319, 184

\bibitem[1991]{Allen91} 
Allen, C., Santill\'an, A.: 1991, \rmxaa, 22, 255

\bibitem[2005]{Amores05} 
Am\^ores, E.B., L\'epine, J.R.D. 2005, \aj, 130, 659

\bibitem[2000]{Ancker00} 
van den Ancker, M.E., Pérez, M.R., de Winter, D., McCollum, B. 2000, \aap, 
363, L25

\bibitem[2001]{Ancker01} 
van den Ancker, M.E., P\'erez, M.R., de Winter, D. 2001, in: {\em Young 
stars near Earth: Progress and Prospects}, eds. R. Jayawardhana, T.P. 
Greene, ASP Conference Series 244, 69

\bibitem[2002]{Andersson02} 
Andersson, B.-G., Idzi, R., Uomoto, A., Wannier, P.G., Chen, B., 
Jorgensen, A.M. 2002, \aj, 124, 2164

\bibitem[1998]{Asiain98}
Asiain, R. 1998, PhD Thesis, Universitat de Barcelona, Spain

\bibitem[1999a]{Asiain99a}
Asiain, R., Figueras, F., Torra, J., Chen, B. 1999a, \aap, 341, 427

\bibitem[1999b]{Asiain99b}
Asiain, R., Figueras, F., Torra, J. 1999b, \aap, 350, 434

\bibitem[2006]{Barrado06}
Barrado y Navascu\'es, D. 2006, \aap, 459, 511

\bibitem[1999]{Barrado99}
Barrado y Navascu\'es, D., Stauffer, J.R., Song, I., Caillaut, J.-P. 1999, 
\apj, 520, L123

\bibitem[2002]{Berghofer02} 
Bergh\"ofer, T.W., Breitschwerdt, D. 2002, \aap, 390, 299

\bibitem[1996]{Bertin96} 
Bertin, G., Lin, C.C. 1996, Spiral Structure in Galaxies: A Density Wave 
Theory, The MIT Press, Cambridge

\bibitem[1991]{Binney91} 
Binney, J., Gerhard, O.E., Stark, A.A., Bally, J., Uchida, K.I. 1991, 
\mnras, 252, 210

\bibitem[1998]{Binney98} 
Binney, J., Merrifield, M. 1998, Galactic Astronomy, Princeton University 
Press, Princeton

\bibitem[1960]{Blaauw60} 
Blaauw, A. 1960, in: {\em Present Problems Concerning the Structure and
Evolution of the Galactic System}, eds. J.H. Oort, H.G. Quik, Nuffic 
Intern. Summer Course, Vol. 3, 1

\bibitem[1964]{Blaauw64} 
Blaauw, A. 1964, \araa, 2, 213

\bibitem[1991]{Blaauw91} 
Blaauw, A. 1991, in: {\em The Physics of Star Formation and Early Stellar
Evolution}, eds. C.J. Lada, N.D. Kylafis, NATO ASI Ser. C, 342, 125

\bibitem[2004]{Breitschwerdt04} 
Breitschwerdt, D., Cox, D.P. 2004, in How does the Galaxy work? A 
Galactic Tertulia with Don Cox and Ron Reynolds, ed. E.J. Alfaro, E. 
P\'erez, \& J. Franco, Astrophysics and Space Science Library, vol. 315, 
391

\bibitem[2006]{Breitschwerdt06} 
Breitschwerdt, D., de Avillez, M.A. 2006, \aap, 452, L1

\bibitem[2003]{Chauvin03}
Chauvin, G., Thomson, M., Dumas, C., Beuzit, J.-L., Lowrance, P., Fusco, 
T., Lagrange, A.-M., Zuckerman, B., Moiullet, D. 2003, \aap, 404, 157

\bibitem[2005]{Clark05} 
Clark, P.C., Bonnell, I.A., Zinnecker, H., Bate, M.R. 2005, \mnras, 359, 
809

\bibitem[1997]{Covino97}
Covino, E., Alcal\'a, J.M., Allain, S., Bouvier, J., Terranegra, L., 
Krautter, J. 1997, \aap, 328, 187

\bibitem[1998]{Cox98} 
Cox, D.P. 1998, in Proceedings of the IAU Colloquium 166: The Local
Bubble and Beyond, ed. D. Breitschwerdt, M.J. Freyberg, \& J. Tr\"umper, 
Lecture Notes in Physics, vol. 506, 121

\bibitem[1987]{Dame87}
Dame, T.M., Ungerechts, H., Cohen, R.S. 1987, \apj, 322, 706

\bibitem[2007]{Dobbs07}
Dobbs, C.L., Bonnell, I.A. 2007, \mnras, 376, 1747

\bibitem[1961]{Eggen61}
Eggen, O.J. 1961, R. Obs. Bull., 41, 245

\bibitem[1965a]{Eggen65a}
Eggen, O.J. 1965a, in: Galactic structure, ed. A. Blaauw, M. Schmidt, 
Univ. Chicago Press, 111

\bibitem[1965b]{Eggen65b}
Eggen, O.J. 1965b, \araa, 3, 235

\bibitem[1998]{Eggen98}
Eggen, O.J. 1998, private communication

\bibitem[1996]{Feigelson96}
Feigelson, E.D. 1996, \apj, 468, 306

\bibitem[2003]{Feigelson03}
Feigelson, E.D, Lawson, W.A., Garmire, G.P. 2003 \apj, 599, 1207

\bibitem[2001]{Fernandez01} 
Fern\'andez, D., Figueras, F. Torra, J. 2001, \aap, 372, 833

\bibitem[2005]{Fernandez05} 
Fern\'andez, D. 2005, PhD Thesis, Universitat de Barcelona, available from 
http://tdx.cesca.es/TDX-0316105-114904/

\bibitem[1998]{Frink98}
Frink, S., Röser, S., Alcal\'a, J.M., Covino, E., Brandner, W. 1998, \aap, 
338, 442

\bibitem[1999]{Frink99}
Frink, S. 1999, PhD Thesis, Universität Heidelberg

\bibitem[2006]{Fuchs06}
Fuchs, B., Breitschwerdt, D., de Avillez, M.A., Dettbarn, C., Flynn, C. 
2006, \mnras, 373, 993

\bibitem[1976]{Georgelin76}
Georgelin, Y.M., Georgelin, Y.P. 1976, \aap, 49, 57

\bibitem[1992]{Geus92}
de Geus, E. 1992, \aap, 262, 258

\bibitem[1989]{Geus89}
de Geus, E., de Zeeuw, P.T., Lub, J. 1989, \aap, 216, 44

\bibitem[2004]{Gittins04}
Gittins, D.M., Clarke, C.J. 2004, \mnras, 349, 909

\bibitem[2002]{Gizis02}
Gizis, J.E. 2002, \apj, 575, 484

\bibitem[1992]{Gregorio-Hetem92}
Gregorio-Hetem, J., L\'epine, J.R.D., Quast, G.R., Torres, C.A.O., de la 
Reza, R. 1992, \aj, 103, 549

\bibitem[2000a]{Hearty00a} 
Hearty, T., Neuh\"auser, R., Stelzer, B., Fern\'andez, M., Alcal\'a, J.M., 
Covino, E., Hambaryan, V. 2000a, \aap, 353, 1044

\bibitem[2000b]{Hearty00b} 
Hearty, T., Fern\'andez, M., Alcal\'a, J.M., Covino, E., Neuh\"auser, R. 
2000b, \aap, 357, 681

\bibitem[1962]{Hoffmeister62}
Hoffmeister, C. 1962, Zeitschr. Astrophys., 55, 290

\bibitem[2001]{Hoogerwerf01}
Hoogerwerf, R., de Brujine, J.H.J., de Zeeuw, P.T. 2001, \aap, 365, 49

\bibitem[2000]{Jayawardhana00}
Jayawardhana, R. 2000, Sci, 288, 64

\bibitem[2005]{Jilinski05}
Jilinski, E., Ortega, V.G., de la Reza, R. 2005, \apj 619, 945

\bibitem[1986]{Kerr86} 
Kerr, F.J., Lynden-Bell, D. 1986, \mnras, 221, 1023

\bibitem[1999]{Knie99} 
Knie, K., Korschinek, G., Faestermann, T., Wallner, C., Scholten, J., 
Hillebrandt, W. 1999, \prl, 81, 18

\bibitem[1973]{Knacke73} 
Knacke, R.F:, Strom, K.M., Strom, S.E., Young, E., Kundel, W. 1973, \apj, 
179,847

\bibitem[1993]{Kurucz93}
Kurucz, R.L. 1993, ATLAS9 Stellar Atmosphere Programs and 2 km/s grid. Kurucz 
CD-ROM No. 13. Cambridge, Massachussets, Smithsonian Astrophysical Observatory

\bibitem[2003]{Lallement03} 
Lallement, R., Welsh, B.Y., Vergely, J.L., Crifo, F., Sfeir, D.M. 2003, 
\aap, 411, 447

\bibitem[1982]{Landau82} 
Landau, L.D., Lifshitz, E.M. 1982, Fluid Mechanics, Pergamon Press, Oxford

\bibitem[2001]{Lawson01} 
Lawson, W.A., Crause, L.A., Mamajek, E.E., Feigelson E.D. 2001, \mnras, 
321, 57

\bibitem[2002]{Lawson02} 
Lawson, W.A., Crause, L.A., Mamajek, E.E., Feigelson E.D. 2002, \mnras, 
329, L29

\bibitem[2007]{Lee07} 
Lee, H.-T., Chen, W.P. 2007, \apj, 657, 884

\bibitem[1999]{Leitherer99} 
Leitherer, C., Schaerer, D., Goldader, J.D., Delgado, R.M.G., Robert, C., 
Kune, D.F., de Mello, D.F., Devost, D., Heckman, T.M. 1999, \apjs, 123, 3

\bibitem[1964]{Lin64}
Lin, C.C., Shu, F.H. 1964, \apj, 140, 646

\bibitem[1969]{Lin69}
Lin, C.C., Yuan, C., Shu, F.H. 1969, \apj, 155, 721

\bibitem[1967]{Lindblad67} 
Lindblad, P.O. 1967, Bull. Astron. Inst. Netherlands, 19, 34

\bibitem[2006]{Lopez06} 
L\'opez-Santiago, J., Montes, D., Crespo-Chac\'on, I., 
Fern\'andez-Figueroa, M.J. 2006, \apj, 643, 1160

\bibitem[1999]{Lowrance99} 
Lowrance, P.J., McCarthy, C., Becklin, E.E., Zuckerman, B., Schneider, G.,
Webb, R.A., Hines, D.C., Kirkpatrick, J.D., Koerner, D.W., Low, F., Meier,
R., Rieke, M., Smith, B.A., Terrile, R.J., Thompson, R.I. 1999, \apj, 512, 
L69

\bibitem[2001]{Luhman01} 
Luhman, K.L. 2001, \apj, 560, 287

\bibitem[2004]{Luhman04} 
Luhman, K.L., Steeghs, D. 2004, \apj, 609, 917

\bibitem[2005]{Luhman05} 
Luhman, K.L., Stauffer, J.R., Mamajek, E.E. 2005, \apj, 628, L69

\bibitem[2002]{Madsen02}
Madsen, S., Dravins, D., Lindegren, L. 2002, \aap, 381, 446 [M02]

\bibitem[2001]{MaizApellaniz01}
Ma\'{\i}z-Apell\'aniz, J. 2001, \apj, 560, 83

\bibitem[2001]{Makarov01}
Makarov, V.V., Fabricius, C. 2001, \aap, 368, 866

\bibitem[2005]{Makarov05}
Makarov, V.V., Gaurne, R.A., Andrievsky, S.M. 2005, \mnras, 362, 1109

\bibitem[2007]{Makarov07}
Makarov, V.V. 2007, \apjs, 169, 105

\bibitem[2005]{Mamajek05}
Mamajek, E.E. 2005, \apj, 634, 1385

\bibitem[2001]{Mamajek01}
Mamajek, E.E., Feigelson, E.D. 2001, in: {\em Young stars near Earth: 
Progress and Prospects}, eds. R. Jayawardhana, T.P. Greene, ASP Conference 
Series 244, 104

\bibitem[1999]{Mamajek99}
Mamajek, E.E., Lawson, W.A., Feigelson, E.D. 1999a, \apj, 516, L77

\bibitem[2000]{Mamajek00}
Mamajek, E.E., Lawson, W.A., Feigelson, E.D. 2000, \apj, 544, 356

\bibitem[2002]{Mamajek02}
Mamajek, E.E., Meyer, E.D., Liebert, J. 2002, \aj, 124, 1670

\bibitem[1981]{Marraco81}
Marraco, H.G., Rydgren, A.E. 1981, \aj, 86, 62

\bibitem[2004]{Merin04}
Mer\'{\i}n, B., Montesinos, B., Eiroa, C. et al. 2004, \aap, 419, 301

\bibitem[1993]{Meynet93}
Meynet, G., Mermilliod, J.-C., Maeder, A. 1993, \aaps, 98, 477

\bibitem[1999]{Mishurov99} 
Mishurov, Yu.N., Zenina, I.A. 1999, \aap, 341, 81

\bibitem[1975]{Miyamoto75}
Miyamoto, M., Nagai, R. 1975, \pasj, 27, 533

\bibitem[1999]{Moreno99}
Moreno, E., Alfaro, E.J., Franco, J. 1999, \apj, 522, 276

\bibitem[1998]{Neuhauser98} 
Neuh\"auser, R., Brandner, W. 1998, \aap, 330, L29

\bibitem[2000]{Neuhauser00} 
Neuh\"auser, R., Walter, F.M., Covino, E. et al. 2000, \aaps, 146, 323

\bibitem[1982]{Olano82}
Olano, C.A. 1982, \aap, 112, 195

\bibitem[2002]{Ortega02}
Ortega, V.G., de la Reza, R., Jilinski, E., Bazzanella, B. 2002, \apj, 
575, L75

\bibitem[2004]{Ortega04}
Ortega, V.G., de la Reza, R., Jilinski, E., Bazzanella, B. 2004, \apj, 
609, 243

\bibitem[2007]{Ortega07}
Ortega, V.G., Jilinski, E., de la Reza, R., Bazzanella, B. 2007, \mnras, 
377, 441

\bibitem[1993]{Palous93} 
Palou\u{s}, J., Jungwiert, B., Kopeck\'y, J. 1993, \aap, 274, 189

\bibitem[1984]{Paresce84} 
Paresce, F. 1984, \aj, 89, 1022

\bibitem[1997]{Poppel97} P\"oppel, W.G.L. 1997, {\em The Gould Belt System
and the Local Interstellar Medium}, Fundamental of Cosmic Physics 18

\bibitem[1999]{Preibisch99} 
Preibisch, T., Zinnecker, H. 1999, \aj, 117, 2381

\bibitem[2002]{Preibisch02} 
Preibisch, T., Brown, A.G.A., Bridges, T., Guenther, E., Zinnecker, H. 
2002, \aj, 124, 404

\bibitem[2001]{Quast01} 
Quast, G.R., Torres, C.A.O., de la Reza, R., da Silva, L., Drake, N.2001, 
in: {\em Young stars near Earth: Progress and Prospects}, eds. R. 
Jayawardhana, T.P. Greene, ASP Conference Series 244, 49

\bibitem[2004]{RanLyo04}
Ran Lyo, A., Lawson, W.A., Feigelson, E.D., Crause L.A. 2004, \mnras, 
347, 246

\bibitem[2003]{Reid03}
Reid, N. 2003, \mnras, 342, 837

\bibitem[2006]{delaReza06}
de la Reza, R., Jilinski, E., Ortega, V.G. 2006, \aj, 131, 2609

\bibitem[1969]{Roberts69}
Roberts Jr., W.W. 1969, \apj, 158, 123

\bibitem[1983]{Rucinski83}
Rucinski, S.M., Krautter, J. 1983, \aap, 121, 217

\bibitem[2001]{Sartori01} 
Sartori, M.J., L\'epine, J.R.D., Dias, W.S. 2001, in: {\em Young stars 
near Earth: Progress and Prospects}, eds. R. Jayawardhana, T.P. Greene, 
ASP Conference Series 244, 98

\bibitem[2003]{Sartori03} 
Sartori, M.J., L\'epine, J.R.D., Dias, W.S. 2003, \aap, 404, 913 [S03]

\bibitem[1999]{Sfeir99} 
Sfeir, D.M., Lallement, R., Crifo, F., Welsh, B.Y. 1999, \aap, 346, 785

\bibitem[1972]{Shu72}
Shu, F.H., Milione, V., Gebel, W., Yuan, C., Goldsmith, D.W., Roberts, 
W.W. 1972, \apj, 173, 557

\bibitem[1984]{Smith84}
Smith, B.A., Terrile, R.J. 1984, Sci, 226, 1421

\bibitem[2001]{Smith01}
Smith, R.K., Cox, D.P. 2001, \apjs, 134, 283

\bibitem[1990]{Snowden90} 
Snowden, S.L., Cox, D.P., McCammon, D., Sanders, W.T. 1990, \apj, 354, 211

\bibitem[1998]{Snowden98}
Snowden, S.L., Egger, R., Finkbeiner, D.P., Freyberg, M.J., Plucinsky, 
P.P. 1998, \apj, 493, 715

\bibitem[1998]{Soderblom98}
Soderblom, D.R., King, J.R., Siess, L. et al. 1998, \apj, 498, 385

\bibitem[2003]{Song03}
Song, I., Zuckerman, B., Bessell, M.S. 2003, \apj, 599, 342

\bibitem[2004]{Song04}
Song, I., Zuckerman, B., Bessell, M.S. 2004, \apj, 600, 1016

\bibitem[1998]{Stauffer98}
Stauffer, J.R., Schultz, G., Kirkpatrick, J.D. 1998, \apjs, 499, 199

\bibitem[2000]{Stelzer00}
Stelzer, B., Neuha\"user, R. 2000, \aap, 361, 581

\bibitem[1995]{Sterzik95}
Sterzik, M., Durisen, R. 1995, \aap, 304, L9

\bibitem[1999]{Sterzik99}
Sterzik, M., Alcal\'a, J.M., Covino, E., Petr, M.G. 1999, \aap, 346, L41

\bibitem[1981]{Straizys81} 
Strai\u zys, V., Kuriliene, G. 1981, \apss, 80, 353

\bibitem[2003]{Straizys03} 
Strai\u zys, V., \u Cernis, K., Barta\u si\=uté, S. 2003, \aap, 405, 585

\bibitem[1984]{Taylor84} 
Taylor, K.N.R., Storey, J.W.V. 1984, \mnras, 209, 5P

\bibitem[1999]{Terranegra99} 
Terranegra, L., Morale, F., Spagna, A., Massone, G., Lattanzi, M.G. 1999, 
\aap, 341, L79

\bibitem[2000]{Torra00} 
Torra, J., Fern\'andez, D., Figueras, F. 2000, \aap, 359, 82

\bibitem[2000]{Torres00}
Torres, C.A.O., da Silva, L., Quast, G., de la Reza, R., Jilinski, E. 
2000, \aj, 120, 1410

\bibitem[2001]{Torres01}
Torres, C.A.O., Quast, G.R., de la Reza, R., da Silva, L., Melo, C.H.F.
2001, in: {\em Young stars near Earth: Progress and Prospects}, eds. R. 
Jayawardhana, T.P. Greene, ASP Conference Series 244, 43

\bibitem[2003]{Torres03}
Torres, G., Guenther, E.W., Marschall, L.A., Neuh\"auser, R., Latham, D.W., 
Stefanik, R.P. 2003, \aj, 125, 825

\bibitem[1998]{Vanhala98}
Vanhala, H.A., Cameron, A.G.W. 1998, \apj, 508, 291

\bibitem[2000]{Voges00}
Voges, W., Aschenbach, B., Boller, T. et al. 2000, Max-Planck-Institut für 
extraterrestrische Physik, Garching, 
http://www.xray.mpg.de/rosat/survey/rass.fsc/

\bibitem[1999]{Webb99}
Webb, R.A., Zuckerman, B., Platais, I., Patience, J., White, R.J., 
Schwartz, M.J., McCarthy, C. 1999, \apj, 512, L63

\bibitem[1998]{Weiler98}
Weiler, K.W., Sramek, R.A. 1998, \araa, 26, 295

\bibitem[2000]{Weinberger00}
Weinberger, A.J., Rich, R.M., Becklin, E.E., Zuckerman, B., Matthews, K. 
2000, \apj, 544, 937

\bibitem[2001]{Weinberger01}
Weinberger, A.J., Becklin, E.E., Zuckerman, B., Schneider, G.,
Silverstone, M.D. 2001, in: {\em Young stars near Earth: Progress and
Prospects}, eds. R. Jayawardhana, T.P. Greene, ASP Conference Series 244,
75

\bibitem[2000]{Weintraub00}
Weintraub, D.A., Kastner, J.H., Hines, D.C., Sahai, R. 2000, \apj, 531, 401

\bibitem[1994]{Welsh94} 
Welsh, B.Y., Craig, N., Vedder, P.W., Vallerga, J.V. 1994, \apj, 437, 638

\bibitem[1999]{Welsh99} 
Welsh, B.Y., Sfeir, D.M., Sirk, M.M., Lallement, R. 1999, \aap, 352, 308

\bibitem[1986]{Wilking86} 
Wilking, B.A., Taylor, K.N.R., Storey, J.W.V. 1986, \aj, 92, 103

\bibitem[1992]{Wilking92} 
Wilking, B.A., Greene, T.P., Lada, C.J., Meyer, M.R., Young, E.T. 1992, 
\apj, 397, 520

\bibitem[1997]{Wilking97} 
Wilking, B.A., McCaughrean, M.J., Burton, M.G., Giblin, T., Rayner, J.T., 
Zinnecker, H. 1997, \aj, 114, 2029

\bibitem[1981]{Yuan81}
Yuan, C., Grosb\o l, P. 1981, \apj, 243, 432

\bibitem[1999]{Zeeuw99}
de Zeeuw, P.T., Hoogerwerf, R., Bruijne, J.H.J., Brown, A.G.A., Blaauw, A. 
1999, \aj, 117, 354 [Z99]

\bibitem[2000]{Zuckerman00}
Zuckerman, B., Webb, R.A. 2000, \apj, 535, 959

\bibitem[2001a]{Zuckerman01a}
Zuckerman, B., Song, I., Bessell, M.S., Webb, R.A. 2001a, \apj, 562, L87

\bibitem[2001b]{Zuckerman01b}
Zuckerman, B., Song, I., Webb, R.A. 2001b, \apj, 559, 388

\bibitem[2001c]{Zuckerman01c}
Zuckerman, B., Webb, R.A., Schwartz, M., Becklin, E.E. 2001c, \apj, 549, 
L233

\bibitem[2004]{Zuckerman04a}
Zuckerman, B., Song, I., Bessell, M.S. 2004, \apj, 613, L65

\bibitem[2004]{Zuckerman04b}
Zuckerman, B., Song, I. 2004, \araa, 42, 685

\end{thebibliography}
\end{document}